\documentclass[10pt,twoside]{article}

\usepackage{amsfonts}
\usepackage{amsmath}
\usepackage{latexsym}
\usepackage{amssymb}
\usepackage[mathcal]{euscript}
\usepackage{amscd}
\usepackage{fancyheadings}
\usepackage{theorem}
\usepackage{citesort}

\def\ben{\begin{equation}}
\def\een{\end{equation}}
\def\bea{\begin{eqnarray}}
\def\eea{\end{eqnarray}}
\def\M{\mathcal{M}}
\def\T{\mathcal{T}}
\def\U{\mathcal{U}}
\def\H{\mathcal{H}}
\def\d{\rm d}
\def\F{\mathcal{F}}
\newcommand{\ket}[1]{| #1 \rangle }
\newcommand{\bra}[1]{\langle #1 |}

\newtheorem{Lem}{Lemma}
\newtheorem{Prop}{Proposition}
\newtheorem{Cor}{Corollary}
\newtheorem{Def}{Definition}
\newtheorem{Thm}{Theorem}
\begin{document}

\title{A group theoretical approach to causal structures \\ and positive energy on spacetimes \vspace{7mm}}
\author{Christophe Patricot \footnote{email: \texttt{cep29@cam.ac.uk}} \vspace{2mm} \\ {\small {\it DAMTP,}}
{\small \it{University of Cambridge,}}  \\
{\small {\it Centre for Mathematical Sciences, Wilberforce Road,}}  \\
{\small {\it Cambridge CB3 0WA, U.K.}}}
\date{}

\maketitle
\thispagestyle{empty}
 \begin{abstract}
This article presents a precise description of the interplay between
the symmetries of a quantum or classical theory with spacetime
interpretation, and  some of its physical properties relating to
causality, horizons and positive energy. 
Our major result is that the existence
of static metrics on spacetimes and
that of positive energy representations of symmetry groups, are 
equivalent to the existence
of particular 
Adjoint-invariant convex cones  in the symmetry
algebras. This can be used to study backgrounds of supergravity and
string theories through their symmetry groups. Our formalism is based
on Segal's approach to
infinitesimal causal 
structures on manifolds.
The Adjoint action in
the symmetry group is shown to correspond to changes of inertial frames
in the spacetime, whereas Adjoint-invariance encodes invariance
under changes of observers. This allows us to give a
group theoretical 
description of the
horizon structure of spacetimes, and also to lift causal
structures to the Hilbert spaces of quantum theories.    
Among other results, by setting up the Dirac procedure for the
complexified universal algebra, we classify the physically 
inequivalent observables of quantum theories. We illustrate this by
finding  
the different Hamiltonians for stationary  observers in $AdS_2$.
  \end{abstract}

\newpage
\pagestyle{fancyplain}
\renewcommand{\sectionmark}[1]{\markboth{\thesection.\,\ #1}{}}
\renewcommand{\subsectionmark}[1]{\markright{\thesubsection\ #1}}
\lhead[\fancyplain{}{\thepage}]{\fancyplain{}{\footnotesize A group
theoretical approach to causal  structures and positive energy on
spacetimes}}
\rhead[\fancyplain{}{\footnotesize \leftmark}]{\fancyplain{}{\thepage}}
\chead[]{}
\cfoot[]{}

\tableofcontents

\section{Introduction} 
The considerable variety of backgrounds allowed in supergravity or
string theories, including spaces with closed timelike curves
\cite{GGHPR,Breckenridge}, or with various types of
singularities
\cite{Breckenridge,Myers,Kallosh1,Kallosh2,PRT},
suggests it would be
useful to analyse
 physically relevant properties of spacetimes  more
 systematically and methodically, in order to obtain
a categorisation of spacetimes as backgrounds hosting physically
different theories. This is further motivated by the difficulty to
reconcile current theories, or even elementary notions of quantum
field theory, with a positive cosmological constant. Part of this
incompatibility results from the fact that supersymmetric bosonic
solutions of supergravity admit a causal Killing field
\cite{GauntlettandPakis}, whereas spacetimes with a positive
cosmological constant seem not to. The difference of impetus
between the $AdS/CFT$ and $dS/CFT$ conjectures also illustrates a
fundamental discrepancy between negatively and positively curved
spaces. Symmetries and supersymmetries of spacetimes
play a decisive r\^ole in determining essential physical
properties such as conservation and boundedness of the energy
\cite{Witten}, stability to perturbations, the existence of
horizons or Ergo-regions, and the existence of closed timelike
curves \cite{GibbonsandHerderio,MaozandSimon,Harris}. In fact we
shall see that many ``physical properties'' of a theory, whether
classical or quantum, such as the
classification of inequivalent Hamiltonians, or the possibility to
define a positive energy, are encoded
directly in its group of symmetries: its structure thus appears to be  as
fundamental as the defining equations of the theory.

Felix Klein, when he
launched the ``Erlanger Programm'' \cite{Klein} in 1872, believed that geometry
itself should consist of the study of properties of a space that
are invariant under a group of transformations. 
The causal structure of a spacetime --by which we mean the transitive
relation between  points which are connected by
timelike or null curves--, encodes some physical properties of
theories built on this spacetime: for example, non-staticity of
spacetime metrics  introduces horizons which
can radiate or more simply laws of physics which appear to be
time-varying. We shall lift the concept of causal structures to the Hilbert
spaces of quantum theories with spacetime interpretation. Klein's motto suggests the causal structures  should
highly depend 
on their groups of symmetries. A well-known example is provided by 
the comparison between  Anti-de-Sitter and de-Sitter
spacetimes. Considered as the homogeneous 
manifolds 
$AdS_{n+1}=SO(2,n)/SO(1,n)$ and $dS_{n+1}=SO(1,n+1)/SO(1,n)$, they have
important structural differences. Philips and Wigner \cite{Wigner3} showed that
all the conjugacy classes of the de Sitter groups 
$O(1,n+1)$ are
ambivalent\footnote{That is, every
element of the group is in the same class as its inverse, or every
vector in the Lie algebra can be transformed to its opposite by
the Adjoint action of a group element.} if $n\equiv 0$ or 3 modulo 4,
while this is not  the case in the
Anti-de-Sitter group $O(2,n)$. We know that $AdS_{n+1}$ admits  a notion  of
positive energy, whereas $dS_{n+1}$ does not. By endowing the symmetry
group of a theory with a causal structure induced from that
of spacetime, we will explain precisely the
interplay between the Adjoint action in the group and spacetime
physics, and in particular explain generalisations of the previous
comment on de-Sitter space. It will  turn out particular
cones in the Lie algebra of the symmetries play a major role, both in
quantum and 
classical theories. The group theoretical formalism we develop, though
more stringent and hence interesting for homogeneous spacetimes, can be
used in any spacetime theory with at least one symmetry. A typical
 set 
of differential equations will often have at least one symmetry, and  hence
our results have a wide application. 

Segal was the first to analyse causal structures on spaces in a
Klein fashion: following Vinberg's work \cite{Vinberg} on convex
cones and obviously motivated by the future cones in General
Relativity,  he defined \cite{Segal} an infinitesimal causal
structure on a manifold to be a smooth assignment of a convex cone
at each point.  He analysed which symmetry groups could act on
such infinitesimal structures and yield (global) causal structures
which have no closed causal curves. Precisely, he focused on the Lie groups
admitting infinitesimal causal structures which are defined by an
Adjoint-invariant cone in their Lie algebra, but did not really
consider homogeneous spaces of such Lie groups. Moreover, Segal
did not justify his assumptions by physical motivations, and it
seems that his followers in this subject, apart from maybe Paneitz
\cite{Paneitz1,Paneitz2}, lead this topic into the
mathematical realm. As a consequence considerable mathematical
progress has been done since the work of Segal, particularly on
the causal structures of symmetric spaces. A recent account of the field is given in
\cite{Olafsson}, and 
other related topics in \cite{Neeb}.
The theory has deep connections with other areas of mathematics
such as semi-groups, hermitian symmetric domains, unitary
representations, many of which commonly have applications in
modern theoretical physics. However, the assumptions of Segal, and
hence most of his results on causal structures, have not been given a
physical sense.  The first aim of
this article is to do so.\\

Starting from general considerations about the symmetries of a
theory with a spacetime interpretation, whether of particles, fields, or
extended objects, we show that these must induce spacetime diffeomorphisms
which preserve the infinitesimal causal structure. The Adjoint
action in the group of symmetries is shown to encode precisely the
effects of changes of inertial frames in the theory. Then we
define observer-independent and static causal structures, formalising
what it means for future-directedness to be
observer-independent. These notions 
highly constrain the action of the group on
spacetime: its Lie algebra must admit an
Adjoint-invariant convex cone, and thus the group  
must satisfy  Segal's
assumptions. All the backgrounds  of supergravity or string theories
which have a causal Killing field satisfy these assumptions. \\
Our second purpose is to classify  the causal
structures of spacetimes admitting Killing symmetries in terms of
properties of their groups of motions. It turns out we can fully
characterise observer-independent causal structures, but also
static metrics, only in terms of particular invariant cones in the
symmetry groups. These essential features are simply encoded
in the Adjoint action of the symmetry group. It becomes clear why
de-Sitter cannot be static, since its group of symmetries does not
admit an Adjoint-invariant cone. These results are directly applicable
to any spacetime with at least one symmetry, and are particularly
simple for product spaces. Our formalism also
allows us to 
deal with non-static spacetimes in an original way: we describe
the horizons of homogeneous spacetimes as horizons in their group
of symmetries, and are then able to see the effect of changes of
inertial frames
and changes of observers on the horizon structure. Horizons are
completely defined group theoretically. \\
Our third purpose is to show that the classical notion of
positive energy --as defined by a causal Killing field on spacetime--,
is related to the
quantum mechanical one --by a one-side bounded operator on a Hilbert
space--. We define a Dirac procedure for the universal algebra of a
symmetry group, and describe the corresponding quantum theory in a
relativistic invariant manner. It turns out the Hilbert space can be
given a causal structure stemming from the locally allowed
Hamiltonians on spacetime, so
that staticity of spacetime metrics is equivalent to the existence of
global time in quantum theory. The physically inequivalent Hamiltonians for
 stationary observers in spacetime are classified as orbits under the
 Adjoint action of the symmetry group. We then use some results of unitary
highest weight representations to show that the existence of an
Adjoint invariant convex cone
in the group of symmetries goes hand in hand with the existence of
positive energy operators for the quantum states. Thus one can
classify symmetry 
groups on 
the physical grounds of staticity of the causal structure,
and the existence of a positive energy. It also follows that for any static
spacetime with a simple group of symmetries, there exists a highest
weight representation, and hence the usual Fock space can be built. \\
 We have tried to make the presentation both
axiomatic and self-contained --which unfortunately is 
incompatible with brevity--, and illustrate our results with examples. The
plan of the paper is as follows.

In Sec.\ref{group actions} we briefly review some salient
properties of group actions on manifolds, and fix some notation
used throughout the article. Our point of view is to describe
Killing vector fields of spacetimes as elements of the Lie algebra
of the group of motions. Sec.\ref{segal causal structures}
introduces infinitesimal causal structures on manifolds and Lie
groups, summarises some of Segal's results \cite{Segal} regarding
bi-invariant convex cones, and raises questions about the physical
interpretation of these results. To answer these, in
Sec.\ref{physical interpretation}, we discuss in detail the link
between the symmetries of a physical theory with a spacetime
interpretation, and the causal structure of the spacetime. We
highlight the difference between changes of inertial frames and
changes of observers, and show that for certain observables,
the Adjoint action in the group of motions represents in the Lie
algebra the effect of changes of inertial frames. Particular
attention is given to the r\^ole of the universal algebra in terms of
defining observables, and to the classification of inequivalent
observables as orbits under the Adjoint action. These
results enable us to give in Sec.\ref{adjoint action and causally
preserving maps} a detailed account of the physical consequences
of the existence of bi-invariant cones in a group of motions. In a
series of Lemmas, we show how these relate to observer-independent
causal structures and static metrics on spacetimes. As an illustration
we give a detailed
analysis of the $AdS_2$ spacetime. Our method provides a classification of 
the inequivalent times of $AdS_2$ in 3 types, which is relevant for the black
hole or matrix model interpretation of this spacetime
\cite{GibbonsandTownsend,Strom1,Strom2}. We then focus
in Sec.\ref{horizons in the group of motions} on homogeneous
spacetimes which admit horizons.  We lift the horizons to loci in the
group itself and describe how and whether some horizons are
related by changes of inertial frames. Then, in Sec.\ref{positive
energy}, we apply the Dirac procedure to the universal algebra, and 
the group structure allows us to define time evolution in a
relativistic invariant way. The notion of observer-dependence allows
us to classify the inequivalent
Hamiltonians of $AdS_2$. We then formally
establish  a relation between the existence of positive energy
operators on a space of states defined by a unitary representation
space of the group of motions, and bi-invariant convex cones in the Lie
group. We conclude
in Sec.\ref{conclusion}.

\section{Some properties of group actions} \label{group actions}
\begin{Def}\label{group action def}
A Lie group $G$ acts on a manifold $\M$ if there exists a smooth
homomorphism from $G$ into the group of diffeomorphisms of $\M$.
In other words there exists a smooth map $\Gamma :\,G \times \M
\to \M$, $(g,p) \mapsto g.p $, such that for all $p\in \M$, for
all $g_1,g_2\in G$, and with $e$ denoting the identity element of
$G$, we have:
 \ben
  e.p=p \quad {\rm {\it and}}  \quad (g_1g_2).p=g_1.(g_2.p) \nonumber
 \een
\end{Def}
The action in this case is called a {\it left} action. For fixed
$p\in \M$ we denote by $\mu_p :G\to \M$, $ g \mapsto g.p$ the
restriction of $\Gamma$ to $G$, and for fixed $g\in G$  by $\nu_g
:\M\to
\M$, $p \mapsto g.p$ the restriction of $\Gamma$ to $\M$. \\
For fixed $p\in \M$ the image
of $\mu_p$ is called the $G$-orbit of $p$, and the group action is said \emph{transitive} if for one (and hence for
all) $p\in \M$, $\mu_p$ is onto. We call $H_p=\{g\in G /\;
\nu_g(p)=p\}$ the stabiliser subgroup of a point $p\in\M$. The
$H_p$ are Lie-subgroups of $G$, and there are all conjugate if the
action is transitive: for all $g\in G$, $H_{g.p}=gH_pg^{-1}$. In
this case, with $H$ the stabiliser group of a chosen point $p\in
\M$, the bijection $gH \mapsto g.p$ between the coset space $G/H$ and
$\M$ (locally compact and connected) can be
made into a diffeomorphism \cite[pp.110-115]{Helgason}
and we call $\M\simeq G/H$ a {\it homogeneous} space. This diffeomorphism chooses
a particular point in $\M$ which corresponds to $eH$.\\
By definition the map $g\mapsto \nu_g$ is a homomorphism of $G$
into the group of diffeomorphisms of $\M$. Its kernel $N$ is a
closed normal subgroup of $G$. If $N=\{e\}$ (or is a discrete
subgroup of $G$) the action is said {\it effective} (or respectively
almost effective). Given $G$ acting on $\M$, the Lie group $G/N$
acts effectively on $\M$, so that there is no restriction in
considering such actions. If $G$ is simple, the action is necessarily almost effective. \\
We denote by $R_g$ and $L_g$ right and left translations in $G$ by
an element $g\in G$, and by $F\ast(x)$ the differential of a
function $F$ at $x$. When convenient we shall represent
 tangent vectors $V_x$ at $x$ by a one-parameter curves $[f(t)]$
such that
$f(0)=x$ and $f'(0)=V_x$. \\
The two following results which  relate the Lie
algebra of a Lie transformation group to special vector fields of the
manifold
 it acts on, are standard, and will be used repeatedly throughout the
 article. They derive
 from Lie's first and second fundamental theorems. \\
 \begin{Thm}
\label{anti-homo} Let $G$ a Lie group with Lie algebra ${\frak g}$
act on a manifold $\M$. Then the map defined by
\begin{eqnarray} \label{killing def}
\phi :  A\in {\frak g} & \longmapsto & \big[ X^A:p\mapsto X^A_p
\equiv \mu_p\ast(e) A = [\exp tA .p ] \big]
\end{eqnarray}
is a Lie algebra anti-homomorphism from ${\frak g}$ into the set
of vector fields of $\M$: $\phi$ is linear and such that
$\phi([A,B])=-[\phi(A),\phi(B)]$, or equivalently
 \ben
X^{[A,B]}=-[X^A,X^B]. \nonumber \een
\end{Thm}
{\it Proof:} For fixed $p\in \M$, $\mu_p$ is a differentiable map
between $G$ and $\M$. For $A\in {\frak g}$ define the right
invariant vector field $\tilde{A}_g=R_g \ast(e)A$ on $G$. Using
$\mu_p\circ R_g=\mu_{g.p}$ together with (\ref{killing def}), we
get
\begin{align*}
\mu_p \ast(g)\tilde{A}_g &=\mu_{g.p}\ast (e)A\equiv X^A_{g.p} \\
  &=X^A_{\mu_p(g)}.
\end{align*}
Thus the vector fields $\tilde{A}$ and $X^A$ are $\mu_p$ -related,
and it follows \cite[p.24]{Helgason} that $\mu_p
\ast(g)[\tilde{A}, \tilde{B}]_g = [X^A, X^B]_{g.p}$. With the
convention that the Lie bracket in ${\frak g}$ is defined as that
of {\it left}-invariant vector fields at the identity, we have $[
\tilde{A}, \tilde{B}]_e = -[A, B]$. Although the $X^A$ are here
defined on the $G$-orbit of $p$ only, this
reasoning can be done
at all $p$, and the theorem follows. $\Box$ \\
For a right action, one gets a homomorphism;
however when
considering isometries of a spacetime, left
actions appear more naturally, so we shall
keep the minus sign. It is clear that in this case $-\phi$ defines a
Lie algebra homomorphism.\\
For $\M$ a connected manifold equipped with a non-degenerate metric tensor
field $h$, we call Killing vector 
   fields of $(\M,h)$ 
 the Killing fields which are complete in $\M$ --so that their flows
  generate one-parameter groups of diffeomorphisms of $\M$--, and
  which close 
  under Lie bracket --so that their flows define a group--. Physically
  this means  that the integral curves of
  the Killing fields should stay in the spacetime, and that the set
  of 
  induced transformations (sometimes called Killing motions) should be
  stable under composition. For maximally
  extended spacetimes, it should not be a problem, and the cases when
  Killing fields are not complete can be attributed to a bad choice of
  coordinates. With this in mind,
  Theorem \ref{anti-homo}  has a converse:
 \begin{Thm} \label{anti-homo converse} Let $\M$ a (connected) manifold
   equipped with a metric tensor field $h$.  Then
 there exists a Lie group $G$ with Lie algebra ${\frak g}$
 which acts on $\M$, such 
that the Killing vector fields of $\M$ are precisely the images of
elements of ${\frak g}$ by the anti-homomorphism defined by
 (\ref{killing def}).
 \end{Thm}
  {\it Proof:} One shows that the transformation group of $\M$ generated by the flows
 of  the Killing fields of $(\M,h)$ can be given a suitable topology
 such that it is Lie group. Its Lie algebra is (anti-)isomorphic to
 that defined by the Killing fields, and this defines a map $\phi$ as
  in (\ref{killing def}). $\Box$ \\
As a consequence, the vector fields $X^A$ in
Theorem \ref{anti-homo} will also be called Killing vector fields,
 though without referring to a particular metric on $\M$.  From now on,
 we shall always think of 
 Killing vector fields of a spacetime $(\M, h)$ as
images of an element of a Lie algebra. 

Theorem \ref{anti-homo} applies to any manifold $\M$, and not just
Lorentzian spacetimes of relativity.  For example, a symmetry group may
act on the 
cotangent bundle of a spacetime (phase-space), or even just on a vector
space of solutions to a differential equation. Even when $\M$ can be
interpreted  as a
 ``spacetime'', it need not be equipped with a
metric such
that $G$-motions are isometries. For
instance the Newton-Cartan 
spacetime \cite{kunzle} and the Newton-Hooke spacetimes
\cite{GibbandPat},  though they admit transitive actions of the Galilei and
Newton-Hooke groups respectively, do not admit invariant
non-degenerate metrics. 
Given any action of a
group on a manifold,  one can
construct  vector fields in $\M$ invariant under
$G$-motion in the sense that $X^A_{g.p}=\mu_{g.p}\ast A=(\mu_p
\circ R_g)\ast A=\mu_p
 \ast (g) \tilde{A}_g$. 
The $\nu_g:p\to g.p$ are mere diffeomorphisms, and the vector fields
$X^A$ their generators. The symmetries become physically
relevant if they correspond to symmetries of particular equations of
motion: the Galilei and Newton-Hooke transforms 
leave invariant respectively Newton's equations in flat space or
their analogue with a cosmological constant term \cite{GibbandPat}. 
 When
one considers  a Lorentzian manifold in Theorem \ref{anti-homo
  converse},  the Killing  motions are
isometries of the metric tensor field, with the additional
physical meaning that the structure of spacetime or more precisely the
geodesic equations  describing the free fall of
particles, do not change along these orbits.  Thus
the diffeomorphisms $\nu_g$  correspond to changes of inertial frames. We shall
 come back to this in detail in Sec.\ref{physical interpretation}.

In the general setting of Theorem \ref{anti-homo}, the properties of the
physics one can do on a homogeneous space $G/H$  depend
on the existence or non-existence of $G$-invariant tensor fields on $G/H$,
which in turn is related to properties of the Adjoint action in
$G$. The latter  
will also play an important r\^ole in  any spacetime $(\M, h)$ with
symmetry group  $G$. 
\begin{Def} Let $G$ a Lie group with Lie algebra ${\frak g}$. The
 Adjoint representation\footnote{In the physics literature, where $G$
   is often a matrix Lie group and hence left and right translations
   are equal to their differentials, one has $Ad_gB=gBg^{-1}$.} of $G$ is
the following group homomorphism:
\begin{align*}
Ad:G &\longrightarrow GL({\frak g}) \\
   g &\longmapsto \big[Ad_g:B\mapsto (L_g \circ R_{g^{-1}})\ast (e)B=
   [g(\exp tB)g^{-1}] \,\big].
\end{align*}
The Adjoint action of a subgroup $H$ of $G$ on ${\frak g}$ is the
group of automorphisms $Ad_h$ for $h \in H$.
\end{Def}
This is not to be confused with the  adjoint representation of ${\frak
  g}$, which the following  Lie algebra
homomorphism: 
\begin{align*}
ad:{\frak g}&\longrightarrow gl({\frak g})  \\
A           &\longmapsto
  \big[\,ad_A:B\to [A,B]\,\big].
\end{align*}
In Lie theory, a homogeneous space $G/H$ of $G$, as opposed to a mere coset
space, is usually defined by a
splitting of ${\frak g}$  into ${\frak g}={\frak h}\oplus {\frak m}$
where ${\frak h}$ is the
Lie algebra of the subgroup $H$ so that $[{\frak h}, {\frak h}]\subset
{\frak h}$,  and where in addition $[{\frak m}, {\frak
    h}]\subset {\frak m}$.  We will call {\it reductive} 
 the homogeneous spaces $G/H$ such that the Lie algebra ${\frak h}$ admits 
an $Ad_H$-invariant complement ${\frak m}$ in ${\frak g}$, so that the
decomposition ${\frak g}={\frak h}\oplus {\frak m}$ is
$Ad_H$-invariant.\footnote{In the general case we just have $Ad_{\exp {\frak
    h}}{\frak m} \subset {\frak m}$ and not $Ad_H{\frak m}\subset
{\frak m}$.} Then the tangent space of $G/H$ at $eH$ is
identified with ${\frak m}$ and the linear isotropy representation of
$H$ with the restriction of the Adjoint action of $H$ on ${\frak
  m}$. It then follows that there is a
one-to-one correspondence between $G$-invariant tensor fields on
$G/H$ (that is, invariant under the $\nu_g$), and $Ad_H$-invariant
tensor fields on ${\frak m}$. For example, this is how one defines
left invariant metrics on homogeneous spacetimes $\M\simeq G/H$ in the
setting of 
Theorem \ref{anti-homo}, so that the physics one can do on such
spacetimes depends on the existence of such $Ad_H$-invariant metrics. 
These facts are
widely known but their details are not
important to us, since we will take a different approach. We will 
consider spacetimes which do not necessarily admit a transitive group
action, and  the Adjoint action will encode properties of their causal
structure.

\section{Segal causal structures and bi-invariant cones}
\label{segal causal structures} We review Segal's approach
\cite{Segal} to causal structures on manifolds and particularly
Lie groups. Generally speaking, the approach lacks precise
physical motivations, and we try to remedy to this first in
Sec.\ref{physical interpretation}. 

A global causal structure on a manifold $\M$ is simply a partial
ordering of its points, ie a transitive antisymmetric relation,
often written $x\prec y$. Antisymmetricity excludes the
possibility of time travel. For example in a time-orientable
spacetime of general relativity one can define: $x \prec y$ if and
only if there exists a curve from $x$ to $y$ whose tangent vector at
every point lies in the cone of future-directed time-like or
light-like vectors.
This is a partial order as long as there are no (non-trivial)
closed timelike or null future-directed curves. The existence of
such a property depends on global features of spacetimes, some of
which are encoded in the symmetries. Segal had the idea of
defining local causal structures, which together with ``well-behaved''
global symmetries, might imply the existence of global causal
structures. With $T_p\M$ denoting the tangent space of $\M$ at $
p$, one  can define the following as in \cite{Segal}:
 \begin{Def} An
infinitesimal causal structure on a manifold $\M$ is a smooth
assignment at every point $p\in \M$ of a non-trivial closed convex cone
$C_p \subset T_p\M$ which is pointed, i.e. such that $C_p\cap -C_p=\{0 \}$. We shall call
$(\M, C_p)$ a Segal structure.
 \end{Def}
We say  a cone $C_p\subset T_p\M$ is {\it Einsteinian} if it
contains a basis of $T_p\M$, or equivalently, if its interior
$Int(C_p)$ (with respect to the topology on $T_p\M$) is non-empty.
Such cones yield the Segal structures of physical interest. Given $(\M,
C_p)$, we will call a curve $t\mapsto \gamma(t)$ in $\M$ {\it
causal} if its tangent vectors at each point lie in the cone at
that point. Of course the reversed curve $t\mapsto \gamma(-t)$ is
{\it not} causal unless $\gamma$ is trivial. Causal vectors and
causal vector fields on $(\M, C_p)$ are defined similarly. Then
the relation $x \prec y$ if there exists a causal curve from $x$
to $y$ defines a transitive relation on $\M$. $\M$ need not be
equipped with a metric --and in fact Segal never does so--
although a sufficiently smooth metric tensor field can provide us
with a Segal structure. This notion can of course be applied to
Lie groups, where in addition some Segal structures can be related
to the group structure: a (non-trivial closed) pointed convex cone
in the Lie algebra ${\frak g}$ of $G$ can be right (or left)
translated to every point in $G$, and hence define a right (or
left) invariant Segal structure $(G,C_g)$, with $C_g=R_g \ast
(e)C_e$ (or $L_g\ast (e)C_e$). Then $C_e \in {\frak g}$ is
invariant under $Ad_G$ if and only if it defines a {\it
bi-invariant} structure, in which case we shall simply denote it
by $(G,C_e)$. It seems natural then to define causally preserving
maps:
 \begin{Def}
 A map between two manifolds equipped with Segal structures is said
  to be causally preserving if
 it sends causal curves to causal
 curves, or equivalently if the
differential of the map sends the cone at each point into the cone
at the image
 point.
 \end{Def}
  One can then consider causally preserving actions of groups on manifolds,
 by requiring that for all $g\in G$, the $\nu_g$ be causally preserving diffeomorphisms
 of $(\M, C_p)$. In the case of homogeneous manifolds, the Segal
 structure on $\M 
 \simeq G/H$  is then
 invariant
 under left translations on cosets, and for reductive spaces, as with
 left-invariant 
 tensor fields,  
  it is uniquely determined by an $Ad_H$-invariant cone in ${\frak m}$ which is
 $\nu_g$-translated in $G/H$.
 We explain in the next section why the symmetries of many physical
 theories induce causally preserving group actions, as is the case  for
 connected groups of motions of
 Lorentzian spacetimes.

For extra mathematical simplicity, one can focus on conal
structures on Lie groups which are both left and right invariant,
and analyse how they map to homogeneous spaces. It will turn out 
that these stringent assumptions  are in fact satisfied as
soon as the group $G$ acts on a static spacetime $(\M, h)$. Paneitz \cite{Paneitz1} introduces bi-invariant
structures by noticing the following: since $G$ and $\M$ can both
be given Segal structures, one can ask that both structures be
related by the full group action, by requiring that $\Gamma$ in
Definition \ref{group action def} be a causally preserving map
from $G \times \M$ to $\M$. We shall call such an action {\it
fully causally preserving}. Explicitly, $\Gamma \ast$
 sends the cones $C_g \oplus C_p$ at $(g,p)$ into the cones
$C_{g.p}$ at $g.p$. We will show in Sec.\ref{adjoint action and
causally preserving maps} that the existence of a bi-invariant
structure on $G$ is necessary  in this case.\\
Segal, Paneitz \cite{Paneitz1,Paneitz2}, Vinberg \cite{Vinberg} and others
 (see \cite{Olafsson}) have classified various Lie groups admitting
bi-invariant convex cones.
 \begin{Thm} \label{Segal thm} \emph{\cite{Segal}} A simple real Lie group $G$ with maximal compact
subgroup $K$ admits an $Ad_G$-invariant (non-trivial closed
pointed Einsteinian) convex cone in ${\frak g}$ if and only if
$G/K$ is a hermitian symmetric space.
 \end{Thm}
{\it Proof:}  This follows from a theorem of  Kostant which states that a real
 finite dimensional vector space $V$ acted
upon by a (connected) semisimple Lie group $G$ with maximal
compact subgroup $K$ admits a non-trivial $G$-action invariant
cone if and only if it admits a non-trivial $K$-action invariant
vector (see \cite{Segal}). $\Box$ \\
 There are four families of irreducible (non-compact) hermitian
 symmetric algebras,  plus
two exceptional ones. Their corresponding simple Lie
groups $G$ and maximal compact subgroups $K$ are given in Table
\ref{irred herm}. 
\begin{table}
\begin{center}
\begin{tabular}{|c|c|}
\hline $G$ & $K$ \\
\hline \hline $Sp(2n,\mathbb{R})$ ($n\geq1$) & $U(n)$ \\
\hline $SU(p,q)$ ($p\geq q\geq 1$) & $S(U(p)\times U(q))$ \\
\hline $SO^{\star}(2n)$ ($n\geq 3$)& $U(n)$ \\
\hline $SO(2,n)$ ($n\geq 3$)& $SO(2)\times SO(n)$ \\
\hline $E_{6(-14)}$ & $SO(2)\times SO(10)$ \\
\hline $E_{7(-25)}$ & $E_{6}\times U(1)$ \\
\hline
\end{tabular}
\end{center}
 \caption{Irreducible hermitian symmetric spaces
$G/K$}
\label{irred herm}
\end{table}
In fact for $G$ simple, $G/K$ is hermitian symmetric if and
only if $K$ has a one-dimensional centre.  The reader is referred to
 \cite[Chapters 8 \& 9]{Helgason} 
  for notations and proofs.  The spaces $G/K$
in Table \ref{irred herm} are also 
 called Cartan classical domains. We shall come back to
these in Sec.\ref{positive energy} through
representation theory,  and their
relevance to positive energy. The exceptional low dimensional
 isomorphisms between some of the groups in this classification imply that some
 of these hermitian spaces are identical. Note also that 
 $\frak{sl}(2, \mathbb{R})\simeq {\frak s}{\frak o}(2,1)\simeq S\frak{p}(2, \mathbb{R}) \simeq \frak{su}(1,1)$, and
 $SO(2,2)/SO(2)\times SO(2)$ is not in the list since it is not
 irreducible but equal to $(SO(2,1)/SO(2))\times (SO(2,1)/SO(2))$.

The simple compact groups are not in this classification, but the
compact groups with non-trivial centres admit Adjoint invariant cones.  
For example, the
set of positive  matrices
(multiplied by $\sqrt{-1}$) defines an Adjoint-invariant pointed cone
in the Lie algebra of the unitary group $U(n)$. In quantum mechanics,
this cone is interpreted as the set of mixed states or their
associated probability functionals \cite{article1}. \\
In the case of semi-simple algebras, one can have direct sums of
invariant cones in the simple summands, with some cones being possibly
trivial, so that one of the summands must be in Table \ref{irred
  herm}. Of course, the cones
might not be pointed nor Einsteinian. The
 Poincar\'e group $E(1,n)$, symmetry group of Minkowski space
 $\mathbb{E}^{1,n}$ equipped with its flat metric, admits a
 bi-invariant cone, but is non-simple. For direct  products of Lie
 groups, if one  group admits an
Adjoint-invariant cone, then so does the product. This will be the
case for the symmetry groups $SO(2,n)\times SO(q)$ of $AdS_{n+1}
\times S^{q-1}$ spacetimes for example. Also, all the
(connected) Lie
groups admitting bi-invariant Lorentzian metrics evidently possess
such cones.  Such groups arise for example 
in the classification of maximally supersymmetric vacua of
supergravity in six dimensions \cite{CFS}. \\

   Paneitz shows in \cite[Theorem 16.5]{Paneitz1}   
that  the universal covers of the classical groups $G$ in
Table \ref{irred herm} admit {\it global} causal structures  which
do not possess closed causal curves: these are defined
by bi-invariant Segal structures $(G, C_e)$ stemming from particular
Adjoint-invariant cones $C_e \subset {\frak g}$. This important
achievement seems 
to be the only a posteriori justification to why bi-invariant cones in
the Lie algebra 
of a group of motion are physically relevant. It is well-known
that closed causal curves can appear when one takes the
quotient of a space, hence this result cannot be extended to all coset
spaces of these groups. Typical examples are provided by 
anti-de-Sitter space or other quotients of its universal cover by
discrete or continuous orbits. More recently, it was shown that the
G\"odel supergravity solution of \cite{GGHPR}, which has closed
time-like curves, 
can be obtained   as a reduction of a six
dimensional plane-wave itself isomorphic to a Lorentzian Lie group
\cite[and references therein]{CFS}. On the other hand, de-Sitter
space equipped with its usual metric does not admit closed causal
curves, though its symmetry group $SO(1,n+1)$ does not admit an
invariant cone. These simple remarks suggest that bi-invariant cones may be
more fundamentally related to staticity of spacetime metrics rather
than to the absence of closed causal curves. Indeed, we will not worry
about the 
global causal structure from now, but explain why Adjoint-invariant cones
are necessary  whenever the notion of future-directedness in a Segal
structure $(\M, C_p)$ is required
to be 
observer-independent.  \\
Before going into the mathematical details, it is necessary to
clarify the physical relevance of group actions on manifolds and
 the r\^ole of the Adjoint  action in the
symmetry groups.

\section{Physical interpretation} \label{physical interpretation}
\subsection{Changes of inertial frames in $\M$}
\label{changes of inertial frames}
 We now justify, from a physical point of view, why it is expected
 that the action  of a
 symmetry group of a theory with spacetime interpretation, induces
 causally preserving diffeomorphisms of an infinitesimal causal
 structure on the spacetime. The space
 of states of the theory is unspecified: it
 is a subset of a real
 vector space, describing particles, fields or extended objects.  Some
properties of the theory, such as notions of what may possibly be
allowed versus what may definitely not, are encoded in a Segal
structure $(\M, C_p)$. This only determines locally at each event
all possible future events, and certainly does not specify the
whole theory. We assume that for all $p\in \M$, any
causal vector $V_p\in C_p$ ``partially describes'' at least one
acceptable state of the theory (for example $V_p$ is the momentum
at $p$ of a one-particle state, or of a 0-brane
\ldots). Mathematically, there exists for each $p\in \M$ a subset 
 $\Omega_p$ of the set of physical states, and a map $\Psi_p$ from
${\Omega}_p$ to the future cone
 $C_p\subset T_p\M$, which is onto. This assumption does violate Heisenberg's
 uncertainty principle.\footnote{We are not saying that the states in
 $\Omega_p$ have position $p\in \M$ and
 momentum $V_p\in C_p$. For example, $\Psi_p$ can be the
 expectation value of the momentum of a wave function at the point
 $p\in \M$.} Moreover, the $\Psi_p$ are assumed to be real linear
 (we have to include mixed states).\\
 The weakest {\it principle of symmetry}
 \cite{Wald} says 
that a symmetry of the theory should take any acceptable state of
the theory to another acceptable state of the theory. We consider here
 {\it active} symmetries, since passive ones, such as gauge
 symmetries, act trivially on the physical states.\footnote{Of course
 gauge theories are included in this discussion, the gauge freedom
 being factored out for clarity. On the other hand, the ``symmetries''
 of string 
 theory which relate physical states of theories with different
 background spacetimes (T-duality for example) are not considered
 here.} A stronger 
principle says that any two states which are related by the
action of a symmetry are {\it physically equivalent} (but
nonetheless different). The latter generally implies that the action
on the states is unitary, but we need not make this assumption here. For our purpose, we shall just assume that
 it takes a state 
``partially described'' by a causal vector at a point, to a different state which is also
``partially described'' by  causal vectors at some other points. Calling  $\Theta(g)$ the action of a particular
 symmetry on the states of the theory, given any $p\in \M$, there exists at least
 one $q\in \M$  such that
 the following diagram holds:
\ben \nonumber 
\begin{array}{ccc}
\Omega_p & \stackrel{\Theta(g)}{\longrightarrow} & \Omega_q   \\
\Big\downarrow\vcenter{\llap{$ \Psi_p \;\; $}}     
&  & \Big\downarrow
\vcenter{\rlap{$ \Psi_q $}} 
\\
C_p &   &  C_q
\end{array}
\een 
We need to specify the physical assumptions further to turn this into
a commutative diagram and show that we can choose $q$ uniquely given $p$.
The  {\it principle of
indistinguishability}, which stems from the principle of relativity,
says that states which are indistinguishable for an observer, get
mapped under a symmetry, to equivalent states which remain
indistinguishable for 
an  equivalent observer. In other words, if
any two
physical states are ``partially described'' by the same vector at $p\in
\M$, they get
mapped by $\Theta(g)$ to physical states such that there exists a
point $q\in \M$ at which they are 
 ``partially described'' by the same vector. Otherwise, $\Theta(g)$
distinguishes the states. If this is not true for a very
general theory with 
 different objects, we can restrict the theory to subsets of states
 such that it become true. Thus for $p\in\M$ fixed, there
 exists $q\in\M$ such that $\Theta(g)({\rm Ker }(\Psi_p))\subset {\rm
   Ker}(\Psi_q)$. Similarly, there exist $r\in \M$ such that
 $\Theta(g^{-1})({\rm Ker }(\Psi_q))\subset {\rm
   Ker}(\Psi_r)$ and thus ${\rm Ker}(\Psi_p)\subset {\rm
   Ker}(\Psi_r)$. The last assumption is that 
 there are sufficiently many physical states to distinguish the points of
 spacetime\footnote{This is related to the existence of (spatially)
   localised states as 
   constructed in \cite{NewtonandWigner2}, which are taken
to physically equivalent ones by $\Theta(g)$. Momentum eigen-states such as
   $e^{-ip_{\mu}x^{\mu}}$ will on the other hand yield $p^{\mu}$
irrespective of the position.}, by which we mean that  
for any two different points $p',r'\in \M$, there exists a vector $v$ (a
real linear combination of physical states),  such that 
$v \in {\rm Ker}(\Psi_{p'})$ but $v \notin {\rm Ker}(\Psi_{r'})$. This
implies that  $p\equiv r$, so that  $\Theta(g)({\rm Ker }(\Psi_p))= {\rm
   Ker}(\Psi_q)$, and hence also that $q\in \M$ is unique given $p$. 
  Thus we can define 
 $\vartheta(g): \M \to \M$, $p\mapsto \vartheta(g)p\equiv q$.  We
 have postulated or physically 
motivated the existence of the following commutative diagram:
 \ben \label{general theory comm diagram}
\begin{CD}
\Omega_p @ > {\Theta(g) } >> \Omega_q   \\
@V{\Psi_p}VV@VV{\Psi_q}V   \\
C_p  @> {\theta(g)} >>  C_q
\end{CD}
\een
which defines the map $\theta(g):C_p \to C_q$.  Since the elements of $C_p$ can
in fact also be interpreted as classical velocities of particular
spacetime observers, we must have, for all $V_p\in C_p$,
$\theta(g)V_p= \vartheta(g)\ast (p)V_p$.  \\
Standard arguments
\cite{Wigner1} show that, if the space of states is a vector
space, the map $g\mapsto \Theta(g)$ defines a representation of
the symmetry group $G$ on the space of states
(possibly after a central extension of $G$ and a modification of
$\Theta$ by a phase \cite{Bargmann}). In all cases  $g\mapsto
\Theta(g)$ is a homomorphism up to factors of indistinguishability
in the space of states. One can then easily derive that for all $g_1,g_2
\in G$, $p\in \M$ and $u \in \Omega_p$,
 \ben
 \theta(g_1)\theta(g_2)\Psi_p(u)=\theta(g_1g_2)\Psi_p(u), \nonumber
 \een
 and the maps $g\mapsto \theta(g)$ and  $g\mapsto \vartheta(g)$ define
 group homomorphisms. If further the symmetry group $G$ is a Lie group, and
 the maps $\Psi_p$ are smooth, letting $\vartheta(g)\equiv \nu_g$ so
 that  $\nu_g \ast \equiv \theta(g)$, we  have the following:
\begin{Lem} \label{general theory lem}
Let $(\M, C_p)$ a Segal structure. Let $G$ a symmetry group of a
physical theory on $\M$ satisfying the assumptions above. Then the
action of $G$ on the physical
 states 
induces, upon restriction to a subset of physical
states, an action on spacetime which sends any causal
vector to a causal vector. There exists a homomorphism $g\mapsto \nu_g$
from $G$ into the diffeomorphisms of $\M$, such that:
 \ben \forall g\in G, \;\; \forall p\in\M, \quad  \nu_g \ast(p)
 (C_p) = C_{g.p}. \label{causally preserving syms}
  \een
\end{Lem}
The induced homomorphism from $G$ into the causally preserving
diffeomorphisms of $(\M, C_p)$ is rarely onto or injective. For
example, the local conformal diffeomorphisms of a Lorentzian
manifold --which are indeed causally preserving-- are seldom all induced by
symmetries of the full physical theory. Typically conformal
invariance is broken in theories with massive particles. On the other hand,
in many theories, multiplication by a global phase factor or
global charge conjugation on the states do not change
velocities so that these symmetries induce the trivial action on
spacetime. The same is true for gauge symmetries. \\
 Of course equation (\ref{causally preserving syms}) follows directly
 if we simply consider
 a theory of causally related events or causal curves on $(\M, C_p)$.
 The explanation given above is far more general.  From now
 on, when we talk of a theory of causal curves, we shall have in mind that it
 is in fact a ``partial description'' of a more general theory satisfying
  the suitable assumptions. Lemma \ref{general theory lem} justifies
 using Segal's  assumption of causally
preserving group actions, on  spacetimes which are not necessarily
homogeneous. 
\begin{Def} \label{ch of inertial frame def} In the setting of Lemma \ref{general theory lem}, or given
  a causally preserving group action on a Segal structure $(\M, C_p)$,  
we call changes of inertial frames or inertial transforms, the spacetime diffeomorphisms
$\nu_g$, for each $g\in G$.   
\end{Def} 
This is a spacetime concept: it may well be that a
symmetry acts non-trivially on the states of the theory, but trivially
on spacetime. The $\nu_g$ relate a frame at $p\in \M$ to a frame
$g.p\in \M$.  In  Minkowski
space,  these generally correspond to the
Poincar\'e transforms in the component of the identity of the full
Poincar\'e group, although some symmetries might be broken or enhanced
in particular theories. The $\nu_g\ast$ are then restricted Lorentz
transforms which relate tangent vectors at different points of
spacetime. Time inversion is never an inertial transform,
since choosing a Segal structure fixes a time 
orientation,  whereas space inversion might be. \\
So far ``inertial'' does not refer to the invariance of
a particular universal law or equation of motion, since we have
simply dealt with causality relations in a manifold.
 We now consider
Lorentzian spacetimes, so that Killing symmetries yield transforms
$\nu_g$ which preserve the geodesic structure of spacetime, and
hence are inertial in the usual relativistic sense.\\

 For completeness we  show that
given a Lorentzian spacetime as in Theorem
\ref{anti-homo converse}, the group $G$ can be taken so that the
$\nu_g$ be causally preserving. This is elementary.  Let $(\M, h)$
a time-oriented manifold  with Lorentzian metric tensor field
$h$, and equip $\M$ with the Segal structure stemming
from $h$ and a time direction. \\
The fact that $X$ is a Killing field of $(\M,h)$ is usually seen
infinitesimally as $\mathcal{L}_{X}\,h=0$: the Lie derivative of
$h$ along the direction of the vector field $X$ vanishes. From a
global point of view, this means that the flow of $X$ is an
isometry: the differential of the map which sends a point $p$ to
another point along an integral curve of $X$ through $p$ preserves
the metric inner product. This flow corresponds to a
diffeomorphism $\nu_g:\M\to\M$ for some $g\in G$, where $G$ is the
Lie group constructed as in Theorem \ref{anti-homo converse} from
the Killing vector fields of $(\M,h)$: indeed, for $A\in {\frak
g}$, the Killing vector field $\phi(A)\equiv X^A$, has flow $\exp
tA.p$. Therefore any Killing motion is equivalent to a global left
action on $\M$ of a group element $g= \textrm{exp}\,tA $, for some
$t\in \mathbb{R}$, and  we can take $G$  connected.
Thus for all $g\in G$, for all $p\in \M$, $\nu_g \ast (p):
T_p\M\to T_{g.p}\M $ is such that:
\begin{equation}
\label{killingisometry} \forall X,Y\,\in T_p\M, \quad
h_p(X,Y)=h_{g.p}(\nu_g \ast (p)X, \, \nu_g \ast (p)Y)
\end{equation}
Equivalently, the pull-back of $h$ under $\nu_g$ is equal to $h$.
This implies the following
\begin{Lem}\label{connected Lie actions}
Let $(\M,h)$ a time-oriented Lorentzian manifold equipped with
its corresponding Segal structure $(\M, C_p)$, and let $G$ a
connected Lie group whose action on $(\M, h)$ represents its
Killing motions. Then for all $g\in G$, the $\nu_g:\M \to \M$, $p
\mapsto g.p$, are causally preserving diffeomorphisms of $(\M,
C_p)$.
\end{Lem}
{\it Proof:} Calling $T$ a (non-vanishing) future directed vector
field in $\M$, define
 \ben
  C'_p=\{ V \in T_p\M \,/\, h_p(V, V) < 0, h_p(V, T_p)< 0 \,
  \}.
 \een
For $t\mapsto g(t)$ a continuous path from $g(0)=e$ to $g(1)=g\in
G$ and $h_p(V, V) < 0$, (\ref{killingisometry}) implies that
$t\mapsto h_{g(t).p} (\nu_{g(t)} \ast (p) V, T_{g(t).p})$ never
vanishes and thus remains strictly negative. Since
$C_p=\overline{C'_p}$, $\nu_g \ast (p) C_p =C_{g.p}$ for all
$g\in G$. $\Box$ \\
 The equivalent of
parallel transport, as an isometry,  along orbits of a group of
motion, is the differential $\nu_g\ast$. More importantly,
$\nu_g\ast(p) X_p$ for an observer at $g.p$ is physically equivalent
to $X_p$ for an observer at $p$: the $\nu_g$ relate
different equivalent states of a theory, one as seen by observers
in a frame at $p\in \M$, to another one as seen by
observers in a frame at $g.p\in \M$. \\

If the symmetries of a physical theory induce via Lemma \ref{general
  theory lem} a transitive action on
the spacetime $\M\simeq G/H$, we get the principal fibre bundle
formulation $(G, H, \M \simeq G/H )$: the isometry group $G$
is the bundle of inertial frames, the fibres $H$ the homogeneous
symmetries which represent inertially related frames at one point, the base manifold  the spacetime $\M$, and the changes of
  inertial 
frames are the left translations in $G$. These act on spacetime as
  well as on the fibers of course. \\
As we will see next, the notion of inertial transforms on  spacetime
  observables  can be described
quite simply and elegantly in the  Lie algebra ${\frak
g}$ of the symmetry group.

\subsection{Changes of inertial frames as Adjoint actions in ${\frak
  g}$ and $\U({\frak g})$}
\label{changes of frames as Adjoint}
 The observables of a physical
theory are usually defined as operators on the space of
states, and generally constitute a real vector space. One also
requires the existence of
real-valued  
functionals associated to each observable, which send a state to the
expectation value of the 
observable on this state. These also constitute a real vector
space. Note that in
quantum theories, real-linearity of the observable functionals (in
their argument) together with the complex vector space structure of
the Hilbert space of pure states,  
requires introducing mixed states.\footnote{In standard notation,
  $[\,\ket{v}\mapsto {\bra v}A{\ket v}\,]$ is neither $\mathbb{C}$ nor
  $\mathbb{R}$-linear in ${\ket v}$,
  whereas 
  $[\,{\ket v}{\bra v}\mapsto {\rm Tr}(A{\ket v}{\bra v})\,]$ is
  $\mathbb{R}$-linear 
in  ${\ket v}{\bra v}$.}  For both mathematical
simplicity and physical relevance, the (operator)
observables must be bounded, or equivalently 
must have a finite operator norm.  The existence of an
associative product law between observables, though refuted
by  many  on physical
grounds (see  \cite{Segalpostulates}), is postulated in
classical mechanics  (Poisson algebras,
see \cite{Arnold} for example), in quantum mechanics \cite{VonNeuman},
quantum 
 field theory, and by extension also in
string theories  (bounded self-adjoint operators). We shall thus
assume 
it. Note that this does
not mean that  the
set of observables is stable under that product. \\ 
In a theory whose symmetry generators constitute a Lie algebra, the {\it universal algebra}\footnote{also
called 
the universal enveloping algebra or the enveloping algebra} of this 
Lie algebra  has the
required structure. 
Moreover, its structure becomes necessary if we make
the following assumption: the generators of the symmetries of a
theory should define
observables whose commutators correspond to the observable
associated to the Lie bracket
of the generators. Then indeed, by definition (see
\cite[p.90]{Helgason}  or 
\cite[Chapter 2]{Dixmier}), the commutator law on the observables is that
stemming from the universal algebra. This is valid for any theory with
symmetries, and not just theories with a spacetime interpretation. We
shall come back  to this in
Sec.\ref{positive energy} with the Dirac quantization procedure. In
this section, we shall motivate the r\^ole of the universal algebra 
 from a spacetime point of view,
by building differential operator observables out of its elements. We will show
that the generalised Adjoint action 
corresponds to the effect of changes of inertial frames on those
particular observables.

For $p\in \M$, we denote by ${\T}({\frak
g})\equiv \mathbb{R}.1 \bigoplus_{n\geq 1} \otimes^n {\frak g}$ and $\T
(T_p\M)\equiv \mathbb{R}.1 \bigoplus_{n\geq 1} \otimes^n T_p\M$ the algebras of
contravariant tensors of any rank on ${\frak g}$ and $T_p\M$
respectively, equipped with tensor multiplication and unit elements $1$.  For $p\in \M$ and
$g\in G$, the linear maps
$\mu_p\ast(e)$, $Ad_g$ and $\nu_g\ast(p)$  
are uniquely extended to morphisms of/between these tensor algebras, using tensor
multiplication. For a
homogeneous manifold, the effect of a change of frame on a
contravariant tensor at a point can be visualised as an Adjoint
action its group of motions:
 \begin{Lem} \label{change of frame and adjoint action}
Let $G$ a Lie group with Lie algebra ${\frak g}$ act transitively
on a spacetime $\M$. Then for all $p\in \M$ fixed,
 \begin{align}
\mu_p\ast( e): \T ({\frak g}) &\longrightarrow \T (T_p\M) \nonumber \\
          T &\longmapsto \mu_p\ast (e) T  \nonumber
 \end{align}
defines a  surjective map. For all $g\in G$, we have the commutative
diagram: \ben \label{adjoint comm diagram}
\begin{CD}
\T ({\frak g}) @ > {Ad_g} >> \T ({\frak g})   \\
@V{\mu_p \ast (e)}VV@VV{\mu_{g.p} \ast (e)} V   \\
\T (T_p \M)  @> {\nu_g \ast (p)} >> \T (T_{g.p} \M)
\end{CD}
\een
 so that the tensor-generalised Adjoint action $Ad_g:\T ({\frak g})\to 
\T({\frak g})$ represents at $e\in G$, the effect of a
 spacetime  inertial transformation
$\nu_g$ on a contravariant tensor at $p\in\M$.
\end{Lem}
 {\it Proof:} That $\mu_p\ast (e)$ is onto follows from the
 fact that it is onto from ${\frak g}$ to $T_p \M$ when $\M \simeq G/H$.
 Let $B\in {\frak g}$. We have $\mu_p\ast (e) B\equiv X^B_p\equiv [\exp tB .p]$ so that
 \begin{align}
  \nu_g\ast (p)\mu_p\ast (e)B &\equiv \nu_g\ast (p)X^B \equiv [g\,\exp tB
.p]=[g\,\exp tB
\,g^{-1}g.p] \nonumber \\
 & =\mu_{g.p}\ast(e) [g\,\exp tB
\,g^{-1}] \nonumber \\
   &\equiv \mu_{g.p}\ast(e) (Ad_gB) \label{adjoint c o f 1}
\end{align}
Diagram (\ref{adjoint comm diagram}) follows by generalisation, with
$Ad_g(A\otimes B)\equiv Ad_g(A)\otimes Ad_g(B)$ and $Ad_g(1)\equiv 1$
and similarly for $\nu_g\ast(p)$. $\Box$
\\

 Diagram
(\ref{adjoint comm diagram}) however is only defined at each fixed point
$p\in \M$, not even on neighbourhoods of a point, and thus is only
relevant for {\it local } effects of changes of inertial
frames. Tensor fields on 
$\M$ do not generally correspond to single elements in the tensor
algebra $\T
({\frak g})$.\\
For a fixed vector  $B\in {\frak g}$, the generalisation of diagram
(\ref{adjoint comm diagram}) to all
$p\in \M$ naturally defines, through equation (\ref{killing def}), the
(Killing) field $X^B$ and we get the following commutative diagram: 
   \ben \label{killing adjoint comm diagram}
\begin{CD}
B\in {\frak g} @ > {Ad_g} >> Ad_gB \in {\frak g}   \\
@V{\phi }VV@VV{\phi} V   \\
X^B  @> {\nu_g \ast } >> X^{Ad_gB}
\end{CD}
\een
 Thus the global effect on a Killing vector field, of a  change of inertial
frame  on spacetime, is particularly
simple:
 \begin{align}
 \forall p\in \M,\quad
&\nu_g\ast(p) X^B_p=X^{Ad_g B}_{g.p} \label{adjoint c o f} \\
{\rm or} \quad [\,p\mapsto X^B_p\,] \;\;&{\rm becomes}\;\;[\,p\mapsto
  X^{Ad_gB}_p\,] \nonumber
\end{align}
We now think of the elements of ${\frak g}$ as defining some observables of a
physical theory. Typically  their associated  Killing fields can be thought
of as differential operators acting on a set of functions on
$\M$. Diagram  (\ref{killing adjoint comm diagram}) shows
that these observables behave simply under inertial transforms, which
suggests they should be useful in the physical interpretation.  We then
define an associative product between these observables, and under the
assumption that the commutator of two observables in a Lie algebra be
given by their Lie  
bracket, we have to consider universal algebras.  We denote by
${\U}({\frak g})$  the universal algebra of
${\frak g}$ and ${\U}(\phi({\frak g}))$ the universal algebra of
the Lie algebra of Killing vector fields on $\M$ defined in Theorem
\ref{anti-homo}.  
\begin{Prop} \label{universal algebra prop} Let $\M$ a spacetime acted
upon almost effectively by its group of symmetries $G$.
With $\widetilde{\phi}$ and $\widetilde{Ad}$ denoting the natural
extensions of the maps $\phi$ and $Ad$ to ${\U}({\frak g})$, we have the
following commutative diagram:
 \ben
\label{universal adjoint comm diagram}
\begin{CD}
 {\U}({\frak g})@ > \widetilde{Ad_g} >> {\U}({\frak g})   \\
@V-\widetilde{\phi }VV@VV-\widetilde{\phi} V   \\
{\U}(\phi({\frak g})) @> \widetilde{\nu_g \ast } >>
{\U}(\phi({\frak g}))
\end{CD}
\een so that the generalised Adjoint action on elements of the
universal algebra ${\U}({\frak g})$ represents the effect of
changes of inertial frames on particular spacetime observables 
on $\M$.\\
Two observables in $\U({\frak g})$ are physically equivalent (for
different observers) if and only if
they are related by an Adjoint action up to scale. 
\end{Prop}
{\it Proof:} This follows from (\ref{killing adjoint comm
diagram}), Theorem \ref{anti-homo}, and the definitions of the
maps involved. Since the action is almost effective, $-\phi$ of
Theorem \ref{anti-homo} defines a Lie
algebra isomorphism between ${\frak g}$ and $\phi({\frak g})$. 
For $A,B \in {\frak g}$
\begin{align}
(-\widetilde{\phi})&\,(A\otimes B-B\otimes A -[A,B])= X^A\otimes
X^B -X^B\otimes X^A -[X^A, X^B] \label{second ideal condition}
\end{align}
 so that elements in the two-sided ideal of $\T ({\frak g})$
 spanned by $A\otimes 
B -B\otimes A -[A,B]$, are mapped into the two-sided ideal of $\T (
\phi({\frak g}))$ spanned by the r.h.s of
(\ref{second ideal condition}).
 Hence $(-\widetilde{\phi})$ is an isomorphism from ${\U}({\frak
 g})$ to ${\U}(\phi({\frak g}))$. \\
Diagram (\ref{killing adjoint comm diagram}) 
extends trivially to the tensor algebras $\T ({\frak g})$ and $\T
(\phi({\frak g}))$, so that, for $A, B \in {\frak g}$ and $g\in
G$, we have:
\begin{align}
\widetilde{\nu_g\ast}\circ (-\widetilde{\phi})&\,(A\otimes
B-B\otimes A -[A,B])  \nonumber \\
&=\widetilde{\nu_g\ast}( X^A\otimes
X^B -X^B\otimes X^A + X^{[A, B]}) \nonumber \\
  &= X^{Ad_gA}\otimes X^{Ad_gB} - X^{Ad_gB}\otimes X^{Ad_gB}
  +X^{Ad_g[A,B]}  \nonumber \\
&= (-\widetilde{\phi})(Ad_gA\otimes Ad_gB -Ad_gB\otimes Ad_gA- Ad_g[A,B]) \nonumber \\
&=(-\widetilde{\phi})\circ \widetilde{Ad_g} \, (A\otimes B-B\otimes A -[A,B])
\nonumber 
\end{align}
which implies that (\ref{killing adjoint comm
 diagram}) extends to diagram (\ref{universal adjoint comm
 diagram}). \\
The last comment follows directly from Definition
 \ref{ch of inertial frame def}. We note again that the physical
 observables will generally correspond to a sub-vector-space of 
 $\U({\frak g})$, which should be Adjoint
 invariant.  $\Box$ \\
Thus the physically inequivalent Killing vector fields of a spacetime
 are classified or labelled
 by the (projective) orbits in ${\frak g}$ under the Adjoint action of
 $G$. However, when one considers a particular observer at $p\in \M$,
 one can only quotient by $Ad_{H_p}$. In Sec.\ref{positive energy}, we
 will show that projective
 orbits in ${\frak g}$ also classify the inequivalent quantum theory
 Hamiltonians of
 stationary observers in $\M$.  We will also 
 study the case of 
 $AdS_2$ in Sec.\ref{AdS2 example}, and show that the classification  
of  the  inequivalent times of this spacetime is clarified in this
 group theoretic  picture.   \\    
In fact,  given any smooth map $U: \M \to {\frak g}$, $[p\mapsto
U(p)]$, the map $[p \mapsto \mu_p\ast (e) U(p)]$ defines
a smooth vector field on $\M$ (and similarly for contravariant
tensors of other ranks). One can easily derive that under a change
of frame $\nu_g$,
 \begin{alignat}{2}
 U:\M &\longrightarrow {\frak g}& \quad {\rm becomes} \quad g(U):\M
 &\longrightarrow  {\frak g}
 \nonumber \\
 p &\longmapsto A(p)&                             p&\longmapsto Ad_g
 A(g^{-1}p) \nonumber
\end{alignat}
and thus represent in ${\frak g}$ the effect of a change of frame
on any vector field on a homogeneous spacetime, and similarly for
contravariant tensor fields. The question is then how big  the set
of relevant observables should be. ${\U}({\frak
 g})$ and ${\U}(\phi({\frak g}))$ are infinite dimensional Lie algebras
 with countable basis (Poincar\'e-Birkhoff-Witt theorem
 \cite[Theorem 2.1.11]{Dixmier}) whereas allowing for any contravariant
 tensor field will generically lead to algebras with
 uncountable basis. We will not consider this further.

It is important to note that, whereas the elements of the tensor
algebra $\T(\phi({\frak g}))$ defined (sums of) contravariant tensor
fields on $\M$ built out of the Killing fields, the elements of the
universal algebra $\U(\phi({\frak g}))$ define particular differential
{\it operators} on $\M$, whose  product is given by composition. 
Proposition \ref{universal algebra prop} shows the relevance of the
universal algebra  of
the symmetry group of a theory, in terms of
particular spacetime observables 
 including the Killing field differential operators on
 $\M$. Precisely, under assumptions we made clear, $\U({\frak g})$
 should contain a minimal set of observables. The spacetime
 need not be homogeneous of course. This also shows that there is no
 need to postulate 
 the effect of changes of inertial frames in the universal algebra:
 the Adjoint action  follows directly from
 the definition of the spacetime changes of inertial frames. From
 this, we will derive in Sec.\ref{Dirac procedure section} the changes of
 inertial frame in the quantum theories. \\

 The results of this subsection, especially Diagram (\ref{killing
 adjoint comm diagram}) 
 and Proposition \ref{universal algebra prop}, give a precise setting
 to address the question of 
the physical meaning of Adjoint-invariant structures in the group
of symmetries of a general theory.  We show  next that,
 whereas the Adjoint action represents at $e\in G$ the effect of
 changes of inertial frames in $\M$, 
 $Ad_G$-invariance is in fact related to invariance under  changes of {\it
observers} in $\M$. For this, we first define what we mean by
changes of observers.

\subsection{Changes of observers}
\label{changes of observers}
 We explained in Sec.\ref{changes of inertial frames} why the action of
 a symmetry group $G$ on a spacetime $\M$ should send causal curves to
 causal curves, and called the induced spacetime diffeomorphisms changes of
 inertial frames. We now ask ourselves
 what it means to require that any {\it fixed} curve which is
 causal for an observer remain causal for distant
 observers.  Generally speaking this question is not well defined: one
 has to compare 
tangent vectors at different points of spacetime, and this requires
 making extra assumptions, since there is no unique way of defining
 common future 
directions at different events. Similar difficulties
 occur when defining  a notion
of simultaneity of events.  One requirement is that the
definition of invariance of ``future-directedness'' under change of
 observers,  must itself be
invariant under the changes of
inertial frames. Moreover, since only observers related by an
inertial transform can observe equivalent states of the theory in
their respective frames, we can only define observer (in)dependence for
 such physically equivalent observers.  It will turn out 
that observer-independent causal structures
require the
existence of causal Killing fields. Although this may suggest
 our approach to causality is too simple, we will show in
 Sec.\ref{positive energy} that it is related to the existence of
 global times and globally valid Hamiltonians in quantum theories,
 which vindicates its physical relevance.

Let $(\M, C_p)$ a Segal structure, and assume for the moment that
the future cones are given by a Lorentzian metric $h$ and a
time-orientation. The symmetries of $\M$ are those of Lemma
\ref{connected Lie actions}. We consider a theory of causal curves or causal
velocities on $(\M, C_p)$,  which we may view as a ``partial description''
of a general theory.  We want to compare a velocity $V_p\in C_p$
which is causal for an  observer at $p\in \M$, to a
possible future direction for an equivalent  observer at $g.p\in \M$.
In a way, we need to postulate the  covariance  rules. The
diffeomorphisms $\nu_g$  map 
$V_p\in C_p$ to $\nu_g\ast (p)V_p \in C_{g.p}$, but up to now this was
interpreted  in
diagram 
(\ref{general theory comm diagram}) as
mapping a state represented by a vector at $p$ to a different but
physically equivalent state represented by a vector at $g.p$. We
 now re-interpret the active transform $V_p \mapsto \nu_g \ast
(p) V_p$ as a 
passive transform, or a change of spacetime coordinates. This is
clear when the group element $g$ is in the stabiliser of $p\in \M$, so
that $g.p$ is a Lorentz boosted or rotated
observer, and we extend it for all $g\in G$. \\
To make the present scenario  clear,
 we define observables 
 with respect to the observers in the spacetime, and dissociate them from the
 states of the 
 theory.\footnote{ In fact we only consider the expectation value of the 
 observables, so that nothing is really measured in the quantum
 sense.}  In a
theory of causal curves in $(\M, h)$, a tangent vector $X_p \in
T_p\M$ defines the functional $V_p \mapsto h_p(X_p, V_p)$, which
we can call a velocity observable for an observer at $p$. $X_p$ is
causal if and only if the expectation value of its observable is
negative for any state with $V_p\in C_p$. Such an observable
measures future directedness at $p$. Reciprocally, a state has a
causal velocity at $p$ if and only if the expectation values of
all causal velocity observables $h_p(X_p, \,.\,)$, with $X_p\in
C_p$, are negative.\footnote{ Equation (\ref{killingisometry}) can be
interpreted either as a change of coordinates on the
representatives of both states and observables, or as an inertial
transform to different but physically equivalent states and
observables.}  \\
Then there are two equivalent ways of defining the question whether a
vector which is causal for an observer at a point, remain causal for
different equivalent observers: one postulates a way
of extending either the states, or the observables, to every
equivalent spacetime point. Proposition \ref{universal algebra prop}
suggests Killing 
fields should be used, when it is possible, to extend our local velocity
observables to every equivalent point.

Consider an observer at the event $p$ in a homogeneous spacetime
$\M\simeq G/H$,  moving
along a causal curve with tangent $X_p$ in the interior of
$C_p$. There exists $A\in {\frak g}$ such that $X_p=X^A_p$, though $A$ is not
generally unique. Since $X^A$
remains causal  in a
neighborhood $\mathcal{N}$ of $p$, the integral curves $\exp tA.
x$ for $x\in \mathcal{N}$ are causal for all $t\in \mathbb{R}$. These
are 
rarely geodesics of $(\M, h)$, but the observers moving along them
are static with respect to each other. (They are called stationary
observers, and will play an important r\^ole in the quantum theories.) Hence for any $A\in {\frak g}$
such that $X^A_p=X_p$, the Killing vector
field $X^A$ is a good candidate for a definition of future
direction extended to a neighbourhood $\mathcal{N}$. In other words
$X^A$ defines 
a Killing observable which extends the velocity observable defined by
$X_p$ to any spacetime point $q \in \M$ by $V_q \mapsto h_q(X^A_q,
V_q)$. Moreover, 
the expectation values 
of this observable are constant along the geodesics of $(\M,h)$ and
hence it yields a constant of motion for states represented by
geodesics. \\
We say that a state $V_p\in C_p$ for an observer at $p\in\M$
with future $X_p\in C_p$ remains causal
for an equivalent observer at $g.p$ if and only if there exists a
Killing observable $X^A$ which extends $X_p$ to $g.p$ and is such that
$V_p$ is causal at $g.p$, in other words: 
\ben
 h_{g.p}( X^A_{g.p}, \nu_g \ast (p)V_p) \leq 0 \label{ch of obs g.p},
  \een
where $V_p \mapsto \nu_g \ast (p) V_p$ is just a change of
coordinates from $p$ to $g.p$.  Requiring
this for all  states $V_p\in
C_p$ and all $g\in G$, given $X_p$ fixed,  
simply implies that $X^A_{g.p} \in C_{g.p}$ for all $g.p\in \M$: 
the observer-independence of the future-directness of states 
is equivalent to the
existence of Killing velocity observables which are causal on a
given $G$-orbit, or equivalently for homogeneous spacetimes, of
 causal Killing fields. One
can say that  such an observable $X^A$ defines a
common clock for the observers on the $G$-orbit of $p\in\M$.\\
One obtains the same conclusion by extending a state with a Killing
field, and letting the observer at $g.p$ measure it with  any local
causal velocity observable $X_{g.p}\in C_{g.p}$. Thus we make the
following    
\begin{Def} \label{observer indep def} Let a Lie group $G$ act on a manifold
$\M$. A Segal structure $(\M, C_p)$ is called observer-independent
if, for any point $p\in \M$ and  any causal vector $X_p\in C_p$, there
exists a (Killing) 
field $X^A$ as defined by (\ref{killing def}) which is causal on
the $G$-orbit of $p$ and such that $X^A_p=X_p$.
\end{Def}
When the Segal structure $(\M, C_p)$ is Einsteinian, its observer-independence
implies the homogeneity of $\M$, since for all $p\in \M$, the $\mu_p\ast
(e)$ must be onto.  Thus
the Killing fields of the definition are causal on the whole
spacetime. Minkowski 
space $\mathbb{E}^{1,n-1}$
 and Anti-de-Sitter space $AdS_{n}$ with their usual metric
 Segal structures, are common examples of such spaces. 
 \begin{Def} \label{static spacetimes def} Let a Lie group $G$ act on
   a manifold 
$\M$. A Segal structure $(\M, C_p)$ is said
 to be static\footnote{In fact, stationary would be more meaningful. Note that
   the Killing fields of Lorentzian spacetimes $(\M,h)$ are assumed to
   have complete orbits.} if there exists a (Killing) field $X^A$ as defined by
 (\ref{killing def}) 
 such that
 for all $p\in \M$, $X^A_p\in Int(C_p)$.
\end{Def}
This definition only makes sense for Einsteinian Segal structures,
which are the physically
relevant cases. Of course, if a Segal structure $(\M,C_p)$ is
observer-independent or static with respect to a certain symmetry
group $G$, it is so for any bigger symmetry group having $G$ as a
subgroup. For Minkowski space, it suffices to consider the abelian
subgroup of the Poincar\'e group generated by the
translations. Note that spacetimes $(\M,h)$ which have a static Killing field
but are not homogeneous cannot have an observer-independent Segal
structure. Four dimensional examples can be found in \cite[Chapter
  16]{Kramer}. There are also examples of spacetimes which admit a
causal Killing field (null in fact), are homogeneous spaces, but whose
causal structure is not observer-independent. The Kaigorodov 
spacetimes \cite{Kaigorodov,Cvetic} and the homogeneous
plane-waves are such 
spaces.  We give
conditions which relate staticity and observer-independence in the
next section. 

Suppose now the group action induces causally preserving diffeomorphisms of
$(\M,C_p)$.  
Clearly equation (\ref{causally preserving syms})  implies that a
Killing field $X^A$ of $(\M, C_p)$ is causal if and 
only if for any $g \in G$, $\nu_g \ast X^A$ is also a causal Killing
  field. Thus Definitions  \ref{observer indep
  def} and \ref{static 
 spacetimes def} are indeed invariant under changes of inertial
 frames. \\
Since the $C_p$ are convex cones, the set of causal Killing fields of
$(\M, C_p)$ is also a convex cone. The Lie algebra of the Killing fields, as a
finite-dimensional vector space, has a standard topology, for which
this cone is closed. Its antecedent under $\phi$ of
(\ref{killing def}) is a closed convex cone in ${\frak g}$. Using diagram
(\ref{killing adjoint comm diagram}), the invariance of the causal Killing
fields under changes of
inertial frames implies that the corresponding cone in ${\frak g}$ is
invariant under the Adjoint action of $G$.  We will be more precise in the
next section. \\

Our discussion of the r\^ole of symmetries in a general physical
 theory  have justified Segal's assumption to consider group actions
 which preserve infinitesimal causal structures.  
We have just shown that observer-independence and staticity of the
 causal 
structures require the existence of  Adjoint invariant
closed convex cones in the Lie algebra of the symmetry group, and
 thus justified physically Segal's second assumption. In fact these
 assumptions hold in a spacetime $(\M,h)$ whenever there is a causal
 Killing field.

\section{Adjoint invariant cones and static spacetimes}
 \label{adjoint action and causally preserving maps}
We precisely formulate observer-independence and staticity of Segal structures
 in terms of properties of the group action. We go from
 the mathematical results to the physical ones, and end the section
 with a detailed discussion on two-dimensional Anti-de-Sitter space.   

\subsection{Fully causally preserving group actions}
 \begin{Lem} \label{causally preserving actions}
 Let $G$ a Lie group with Lie algebra ${\frak g}$ act on manifold $\M$ non-trivially
 but not necessarily transitively. If there exists
 Segal structures $(G, C_g)$ and $(\M, C_p)$ such that the action
 $\Gamma: G \times \M \to \M $ is fully causally preserving, then there exists a
 non-trivial (proper) closed convex cone $\widetilde{C}_e\subset {\frak g}$
 which is stable under $Ad_G$. If the action is almost effective, $\widetilde{C}_e
 \cap-\widetilde{C}_e=\{0\}$, and further if $G$ is simple, $C_e$ is Einsteinian.
\end{Lem}
{\it Proof:} If $\Gamma$ is causally preserving, then so are
$\mu_p: G \to \M$ and $\nu_g:\M \to \M$ for all $p\in \M$ and
$g\in G$. Thus for each $p\in \M$, $C_e^p \equiv \{ A \in {\frak
g} / \mu_p \ast (e) A \in C_p \}$ is a closed convex cone
containing $C_e$.  The $\nu_g$ are causally preserving
diffeomorphisms of $\M$:
\begin{align*}
V_p\in C_p &\Leftrightarrow \nu_g \ast (p) V_p \in C_{g.p}
\end{align*}
so that using (\ref{adjoint c o f}) we have:
\begin{align}
 A \in C_e^p &\Leftrightarrow X^A_p \ C_p  \nonumber \\
     &\Leftrightarrow \nu_g\ast (p)X^A_p = X^{Ad_gA}_{g.p} \in C_{g.p}
     \nonumber \\
     &\Leftrightarrow Ad_gA \in C_e^{g.p} \label{adjoint related cones 3}
\end{align}
In other words $C_e^{g.p}=Ad_g C_e^p$. Define:
 \begin{align}
 \widetilde{C}_{e} &\equiv \bigcap_{p\in \M} C_e^{p}\equiv \bigcap_{p\in \M}
  \{ A \in {\frak g}\, / \,\mu_p \ast (e) A \in C_p \} \nonumber \\
 \label{adjoint invariant cone def}  &\equiv \{ A \in {\frak g}
 \, / \, \forall p\in \M, \; X^A_p \in C_p \}
\end{align}
so that for all $g \in
 G$,
 \ben
 \widetilde{C}_e = \bigcap_{ p\in
 \M} C_e^{g.p} = \bigcap_{p\in \M} Ad_gC_e^p = Ad_g
 \widetilde{C}_e\,. \nonumber
 \een
 $\widetilde{C}_{e}$ is a closed convex cone in
 ${\frak g}$ containing $C_e$, and it is invariant under
 $Ad_G$. In addition, we have:
 \ben
  \widetilde{C}_{e}\cap-\widetilde{C}_{e} = \bigcap_{p \in \M}
  (C_e^p \cap -C_e^p)= \bigcap_{p\in \M} \textrm{Ker}(\mu_p \ast (e)),
  \label{kernel intersection}
  \een
and $\widetilde{C}_{e}\cap-\widetilde{C}_{e} \neq {\frak g}$ since
there exist $p\in \M$ such that $\mu_p$ is not trivial. Hence
$\widetilde{C}_{e}$ is a non-trivial proper subcone of ${\frak
g}$. Moreover, if the action is almost effective, the map $\phi$
in Theorem \ref{anti-homo} is injective, so (\ref{kernel
intersection}) implies that $\widetilde{C}_e$ is pointed. In all
cases, $\widetilde{C}_{e}\cap -\widetilde{C}_{e}$ is a
 vector space in ${\frak g}$ which is $Ad_G$-invariant, so it is
 an ideal of ${\frak g}$. The same is true for the vector span of
 $\widetilde{C}_e$  in ${\frak g}$.
 Thus if ${\frak g}$ simple, $\widetilde{C}_{e}$ is pointed and
 Einsteinian. $\Box$  \\
Note that if the cones $C_p$ are not pointed, as in a Newtonian
theory on $\M$, (\ref{kernel intersection}) does not hold and
$\widetilde{C}_e$ is not necessarily pointed (but it is still
(pointed) Einsteinian whenever $G$ is simple).
 Lemma
\ref{causally preserving actions} has the following converse:
\begin{Lem} \label{causally preserving converse} Let $G$ a Lie group
  admitting a non-trivial 
$Ad_G$-invariant pointed closed
 convex cone $C_e \subset {\frak g}$, and  let $G$ act transitively on
 a manifold $\M \simeq G/H$. Suppose 
there exists $p\in \M$ such that the closure of $\mu_p\ast(e)C_e$ in
$T_p\M$ is a  (non trivial) pointed cone.  Then $G$ and $\M$
can be given Segal structures such that the action $\Gamma: G
\times \M \to \M$ is fully causally preserving. \\
Calling ${\frak h}$ the Lie
algebra of $H=H_p$, if ${\frak g}= {\frak h}\oplus {\frak m}$ is an
$Ad_H$-invariant split of ${\frak g}$, this condition is equivalent to
$\pi_{\frak m}(C_e)$ is pointed and closed in ${\frak m}$, where
$\pi_{{\frak m}}$  the projection of
${\frak g}$ on ${\frak m}$. 
\end{Lem}
{\it Proof:} Let $C_g= R_g \ast C_e =L_g \ast Ad_{g^{-1}} C_e=L_g
\ast C_e$ define a (bi-invariant) Segal structure on $G$. Call
$C_p\equiv \mu_p\ast(e)C_e\equiv \{ X^A_p \;/\;A\in C_e\}$. We
first show that the map $g.p \mapsto  \nu_g\ast(p)C_p \equiv C_{g.p}$
defines, up to closure, a Segal
structure on $\M$. \\
For $h\in G$ such that $hg.p=g.p$,  using (\ref{adjoint c o f})
and $Ad_G$-invariance of $C_e$, we have:
\begin{align}
 C_{g.p}&\equiv  \nu_g\ast(p)C_p = \{ X^{Ad_gA}_{g.p} \; /\; A \in C_e \}=\{ X^{B}_{g.p} \; /\; B \in C_e
 \} \label{cone field} \\
\textrm{so that} \quad
 C_{hg.p}&= \{ X^{Ad_hB}_{hg.p} \; /\; B \in C_e \}= \{
 X^{Ad_gA}_{g.p} \; /\; A \in C_e \} \nonumber \\
       &= C_{g.p} \nonumber 
 \end{align}
Since the $\nu_g$ are diffeomorphisms and $\overline{C_p}$ is
pointed by hypothesis,
$g.p\mapsto \overline{C_{g.p}}$ is a smooth assignment on $\M$ of
(closed) pointed cones; hence $(\M, \overline{C_{g.p}})$ is a Segal
structure.  \\
By definition the $\nu_g$ are causally preserving. For all $p\in
\M$ and $g\in G$, using $\mu_p \circ R_g= \mu_{g.p}$ and (\ref{cone field}), 
 \begin{align}
\mu_p\ast (g) C_g &= \mu_p \ast (g) R_g \ast (e) = \mu_{g.p} \ast
(e) C_e  \nonumber \\
\label{mu p caus pres}   &= \{ X^{A}_{g.p} \; /\; A \in C_e \}=
C_{g.p} \subset \overline{C_{g.p}}
\end{align}
so that the $\mu_p$ are causally preserving. Now for any $(g,q)\in
G \times \M$, as $C_g=L_g \ast C_e$ and $C_{q}= \mu_{q}\ast(e) C_e$,
any element of $C_g\oplus C_{q}$ is represented by a curve
$[(g\exp tB, \exp tA.p)]$ with $A,B \in C_e$, and
\begin{align*}
\Gamma \ast (g,p)[(g\exp tB, \exp tA.p)] &\equiv [\Gamma (g\exp
 tB,
\exp tA. p)] = [(g\exp tB \exp tA).p] \\
   & = [ \mu_p (g\exp tB \exp tA) ] = \mu_p\ast (g) [g \exp tB \exp
   tA]
\end{align*}
From (\ref{mu p caus pres}) it suffices to show that $[g \exp tB
\exp tA]\in C_g$ or equivalently that $[\exp tB \exp tA] \in C_e$.
This follows from (see \cite{Helgason} p.96 for example)
 \ben
\exp tB \exp tA  =  \exp( t(B+A) + {t^2 \over 2}[B,A] + O(t^3))
\nonumber  \een which implies that $[\exp tB \exp tA] = B+A \in
C_e $. By continuity we get $\Gamma(C_g \oplus
\overline{C_{q}})\subset\overline{C_{q}}$,  hence
 $\Gamma$ is causally preserving. \\
Taking $H=H_p$ the stabiliser subgroup of $p\in \M\simeq G/H$,
${\frak m}$ is isomorphic to the tangent space of $G/H$ at $eH$,
so to $T_p\M$, under:
\begin{align}
{\frak m} &\to T_{eH}G/H     \nonumber \\
\label{tangent space diffeo}    A &\mapsto [\exp tA \, eH]
\end{align}
Via the diffeomorphism $G/H \simeq \M$, (\ref{tangent space
diffeo}) corresponds to the restriction to ${\frak m}$ of
$\mu_p\ast (e): {\frak g} \to T_p\M$. This restriction is then an
isomorphism, ie ${\frak h}= {\rm Ker} (\mu_p\ast (e))$, and
$\mu_p\ast (e) C_e = \mu_p\ast (e) (\pi_{\frak m}(C_e))$ is
pointed and closed if and only if $\pi_{\frak m}(C_e)$ is pointed
and closed in the topology of ${\frak m}$.
 $\Box$ \\
Note that if in addition $C_e$ is Einsteinian, $\mu_p\ast (e)C_e$
spans $T_p\M$ so the Segal structure defined on $\M$ is Einsteinian.

\subsection{Observer-independent and static causal structures}
We now relate the two previous Lemmas on fully preserving group
actions  to the physical formalism
of Sec.\ref{physical interpretation}. All the results derive quite
simply, and some may seem to
be just restatements for the thorough reader. We show how
observer-independence and staticity of spacetime causal structures
drastically reduce the variety of allowed spacetime symmetry groups. We
hope to make  the physical relevance clear.
\subsubsection{Observer-independent causal structures} 
\begin{Lem} \label{causal killing field lem} Let $G$ a Lie group act
  almost effectively  on a manifold $\M$ with
a Segal structure $(\M, C_p)$. Suppose that for all $g\in G$, the
$\nu_g$
 are causally preserving diffeomorphisms of $(\M,
C_p)$. Then $(\M, C_p)$ admits a causal (Killing) field as defined
by (\ref{killing def}) if and only if there exists a bi-invariant
Segal structure $(G, C_e)$ such that the action $\Gamma: G\times
\M \to \M$ is fully causally preserving. 
\end{Lem}
{\it Proof:} $[\Rightarrow]$ As in the proof of Lemma
\ref{causally preserving actions}, $C_e \equiv \{A\in {\frak g}
 \,/\,\forall p\in \M, X^A_p\in C_p \}$
  defines an $Ad_G$-invariant pointed closed convex cone. It is non-empty
 by hypothesis, and we can define the bi-invariant Segal structure $(G, C_e)$.
  Using the end of the proof of Lemma \ref{causally preserving
  converse}, we show that $\Gamma: G \times \M \to \M$ is fully
  causally preserving.

$[\Leftarrow]$ Let $A\in C_e$, so that for all $p\in \M$,
$X^A_p\equiv \mu_p\ast (e)A\in C_p$ by hypothesis. Since the
action is almost effective, $X^A$ is non-trivial. $\Box$ \\
When $\M$ is homogeneous, we just need a condition at a point:
\begin{Cor} \label{causal killing field cor}
Let $(\M,C_p)$ and $G$ as in Lemma \ref{causal killing field lem}.
If $\M$ is homogeneous, then $(\M, C_p)$ admits a causal (Killing)
field  if and only if there exists an Adjoint-invariant (closed
pointed convex) cone $C_e\subset {\frak g}$ and a point $p\in \M$
such that $\mu_p\ast(e) C_e \subset C_p$ and is non-trivial.
\end{Cor}
We can now straightforwardly characterise observer-independent
Segal structures:
\begin{Prop} \label{obs independent prop} Let $G$ a Lie group act
  almost effectively and transitively on a manifold $\M$ 
with Segal structure $(\M, C_p)$. Suppose that for all $g\in G$,
the $\nu_g$ are causally preserving diffeomorphisms of $(\M,
C_p)$.
 Then $(\M, C_p)$ is observer
independent if and only if the following conditions are
satisfied:\\
(i) there exists a bi-invariant Segal structure $(G, C_e)$ such
that action $\Gamma: G \times \M \to \M$ is fully causally preserving, \\
(ii) there exists $p\in \M$ such that $\mu_p\ast(e)C_e=C_p$.
\end{Prop}
{\it Proof:}
 $[\Rightarrow]$ {\it (i)} is the same as in Lemma \ref{causal killing field
 lem}, so that {\it (ii)} follows by Definition \ref{observer indep
 def}.

$[\Leftarrow]$ Let $q\in \M$, $V_q\in C_q$. There exist $g\in G$
and $V_p \in C_p$ such that $g.p=q$ and $\nu_g\ast(p)V_p=V_q$.
Then there exists $A\in C_e$ such that $X^A_p=V_p$ extends $V_p$
to a causal (Killing) field. Then $X^{Ad_gA}_q$ extends $V_q$ to a
causal (Killing) field. $\Box$ \\
The transitivity condition is superfluous when $(\M, C_p)$ is
Einsteinian. The following version of Proposition \ref{static
spacetimes prop} will be more useful (see Sec.\ref{AdS2 example}):
\begin{Cor} \label{obs indep cor}
Let $(\M, h)$ a time-oriented Lorentzian spacetime, and let $G$ a
(connected) Lie group with Lie algebra ${\frak g}$ represent the
Killing motions of $(\M, h)$ as in Lemma \ref{connected Lie
actions}. Then the metric Segal structure $(\M, C_p)$ is observer
independent if and only
if the following conditions are satisfied: \\
(i) there exists $C_e$ a pointed closed convex cone in ${\frak g}$
which is $Ad_G$-invariant, \\
(ii) there exists $p\in \M$, $\mu_p\ast(e)C_e=C_p$.
\end{Cor}

\subsubsection{Static spacetimes}
From now the spacetime Segal structures $(\M, C_p)$ are Einsteinian so
that Definition \ref{static spacetimes def} makes sense.
\begin{Prop}\label{static spacetimes prop}
Let $G$ a Lie group act almost effectively on a manifold $\M$ with
a  Segal structure $(\M, C_p)$. Suppose that for all
$g\in G$, the $\nu_g$
 are causally preserving diffeomorphisms of $(\M,
C_p)$.
 Then $(\M, C_p)$ is static if and only if the following conditions are
satisfied:\\
(i) there exists a bi-invariant Segal structure $(G, C_e)$ such
that action $\Gamma: G \times \M \to \M$ is fully causally preserving, \\
(ii) $C_e \cap \big( \bigcap_{p\in \M}\{ A\in {\frak g} \,/\,
X^A_p \in Int(C_p)\}\big)\neq \emptyset $.
 \end{Prop}
{\it Proof:}  This is straightforward since $\bigcap_{p\in \M}\{
A\in {\frak g} \,/\, X^A_p \in Int(C_p)\} \neq \emptyset$ is just
Definition \ref{static spacetimes def} \\
 $[\Rightarrow]$ $C_e
\equiv \{A\in {\frak g}
 \,/\,\forall p\in \M, X^A_p\in C_p \}$
  defines a non-emtpy $Ad_G$-invariant pointed closed convex cone.
  Using the end of the proof of Lemma \ref{causally preserving
  converse}, we show that $\Gamma: G \times \M \to \M$ is fully
  causally preserving.\\
$[\Leftarrow]$ Any element $B\in C_e \cap \big( \bigcap_{p\in
\M}\{ A\in {\frak g} \,/\, X^A_p \in Int(C_p)\}\big)$ defines a
causal (Killing) vector field $X^B$ on $(\M, C_p)$. $\Box$ \\
$SO(1,n)^+$ --the identity component of $SO(1,n)$--, as we noticed
earlier, does  not admit an Adjoint
invariant cone. However by Lemma \ref{connected Lie actions} it
acts by causally preserving diffeomorphisms  on de Sitter space
$dS_n=SO(1,n)^+/SO(1, n-1)^+$ equipped with its usual Einstein
metric. Propositions \ref{obs independent prop} and \ref{static
  spacetimes prop} then imply that the causal structure of de-Sitter
space can be  neither
observer-independent  nor static.\\
In contrast with de-Sitter space, there can exist bi-invariant cones
in the group of motions which do not satisfy all the requirements
for staticity or the existence of a causal Killing field. The fact
that the action $\Gamma$ be fully causally preserving in Lemma
\ref{causal killing field lem} and Proposition \ref{static
spacetimes prop} is essential. Consider the Schwarzschild metric:
 \ben
 {\d} s^2= -\bigg(1-{2M \over r}\bigg){\d} t^2 + \bigg(1-{2M \over
 r}\bigg)^{-1} {\d} r^2 + r^2( {\d} \theta ^2+ \sin ^2 \theta {\d} \psi
 ^2) \label{B-H metric}
  \een
It has symmetry group $\mathbb{R} \times SO(3)$, with $\partial_t$
generating the commuting $\mathbb{R}$ symmetry. Thus the element $T\in
\mathbb{R}T\oplus {\frak so}(3)$ such that $\partial_t\equiv X^T$,
is stable under the Adjoint action of $\mathbb{R} \times SO(3)$.
The invariant cone $\{\lambda T / \lambda \geq 0\}$ is mapped into
the future cones $C_p$ at $p\in \M$ only when $r\geq 2M$. The
group action $\Gamma$, though it is fully causally preserving on the
region $r\geq 2M$, is not fully causally preserving on
the whole of spacetime. Since inside the horizon all the
Killing vectors become spacelike, clearly $ \bigcap_{p\in \M}\{ A\in
{\frak g} \,/\, 
X^A_p \in C_p\}= \emptyset$, and the Schwarzschild spacetime is not
static.  This example also shows that Corollary \ref{causal killing field
cor} is not true when the spacetime is not homogeneous. The following
conditions for staticity will be more useful: 
 
\begin{Cor} Let $G$ a Lie group act transitively on a manifold
  $\M$, and suppose that for all $g\in G$, the $\nu_g$ are causally
  preserving diffeomorphisms of the Segal structure
  $(\M, C_p)$. If the 
Adjoint invariant cone $C_e=\bigcap_{p\in \M}\{ A\in {\frak g}
\,/\, X^A_p \in C_p\}$ has non-empty interior in ${\frak g}$, then $(\M,
C_p)$ is static. Furthermore, if $G$ is simple, then $(\M, C_p)$ is static if
  and only if  $C_e\neq \{ 0\}$. Thus if $(\M, C_p)$
 is observer-independent and $G$ is simple, $(\M, C_p)$ is static. 
\end{Cor}
{\it Proof:} We need to show that $Int(C_e)\neq \emptyset$ implies
{\it (ii)} of Proposition \ref{static spacetimes prop}, since {\it
  (i)} follows immediately. Let $A\in Int(C_e)$, so that there exists
an open neighborhood of the origin $\mathcal{N}$ such that
$A+\mathcal{N} \subset C_e$. Let $p\in \M$. By definition of $C_e$,
$X^A_p + \mu_p\ast(e)(\mathcal{N})\subset C_p$. Since the action is
transitive, $\mu_p\ast(e):{\frak g}\to T_p\M$ is onto. Let ${\frak m}_p$ such that ${\frak g}={\rm
  Ker}(\mu_p\ast(e))\oplus {\frak m}_p$, so that $\mu_p\ast(e):{\frak
  m}_p \to T_p\M$ is an isomorphism and hence an open mapping for the
induced topology on ${\frak m}_p$. It follows that
$\mu_p\ast(e)(\mathcal{N}\cap{\frak m}_p)$ is an open in $T_p\M$, so
that $X^A_p+ \mu_p\ast(e)(\mathcal{N}\cap{\frak m}_p)$ is a
neighborhood of $X^A_p$ contained in $C_p$, hence $X^A_p \in
Int(C_p)$.\\
When $G$ is simple, $C_e=\{0\}$ or $Int(C_e)\neq \emptyset $, thus
staticity of 
$(\M, C_p)$  is equivalent to $C_e\neq \{0\}$. If $(\M, C_p)$ is
observer-independent, $C_e$ in Proposition \ref{obs independent prop}
is such that $\{0\}\neq C_e \subset \bigcap_{p\in \M}\{ A\in {\frak g}
\,/\, X^A_p \in C_p\}$.  $\Box$\\  
\begin{Cor} Let $(\M, h)$ a time-oriented Lorentzian spacetime with
  Segal structure $(\M, C_p)$, and
  let $G$ a Lie group act on $\M$ as in Lemma \ref{connected Lie
  actions}. Suppose  that for each $p\in \M$, the $G$-orbit of $p$ has
  dimension 2 at least. If the Adjoint-invariant cone
  $C_e\equiv\bigcap_{p\in \M}\{ A\in {\frak g} 
 \,/\, X^A_p \in C_p\}$ has non-empty interior, then $(\M, C_p)$ is
  static.
\end{Cor}
{\it Proof:} We follow the previous proof, and if $A\in Int(C_e)$,
there is a ``neighborhood'' of $X^A_p$ which is at least
two-dimensional and sits in $C_p$, so
that $X^A_p$ cannot be light-like. $\Box$  \\  
We shall see on an example in the next subsection that we can have $A\in
C_e\backslash Int(C_e)$ and yet $X^A$ static. \\

We now relate staticity
to observer-independence of the causal structure. 
For a Segal structure $(\M, C_p)$ with causally preserving
diffeomorphisms $\nu_g$, we say a cone $C_p\subset T_p\M$ is {\it
homogeneous} if the action of the $\nu_h\ast(p)$ for $h \in H_p$
on $T_p\M$ is transitive on the rays of the interior of $C_p$. In
a spacetime $(\M, h)$, the $\nu_h$ generate a subgroup of the
restricted Lorentz group $SO(1, n-1)^+$. 
\begin{Lem} Let $(\M, C_p)$ a  Segal structure, and let $G$ act
on $\M$ such that for all $g\in G$, the $\nu_g$  are causally
preserving diffeomorphisms of $(\M, C_p)$. Suppose $(\M, C_p)$ is
static, so that $C_e=\bigcap_{p\in \M}\{ A\in {\frak g} \,/\,
X^A_p \in C_p\}$ is a non-trivial $Ad_G$-invariant closed convex
cone in ${\frak g}$. If there exists $p\in \M$ such that $C_p$ is
a homogeneous cone, and $\mu_p\ast(e)C_e$ is closed in $T_p\M$,
then $(\M, C_p)$ is observer-independent. 
\end{Lem}
{\it Proof:} Let $X^A$ a static Killing field so that $A\in C_e$,
and let $V_p\in Int(C_p)$. By hypothesis there exists $h\in H_p$
and $\lambda > 0$ such that $V_p = \nu_h \ast (p)(\lambda X^A)=
X^{Ad_h(\lambda A)}_p$, using (\ref{adjoint c o f}). The Killing
field $X^{Ad_h(\lambda A)}$ is an extension of $V_p$ which is
causal since $Ad_h A \in C_e$. Thus $Int(C_p) \subset
\mu_p\ast(e)C_e\subset C_p$, so that $\mu_p\ast(e)C_e=C_p$ if
$\mu_p\ast(e)C_e$ is closed. The action $\Gamma$ is fully causally
preserving by Proposition \ref{static spacetimes prop}, and $\M$
is necessarily homogeneous,
so  we can apply Proposition \ref{obs independent prop}. $\Box$ \\
For example, the staticity of Anti-de-Sitter space and Minkowski space
implies the 
observer-independence of their causal structures. 

\subsection{The inequivalent times of $AdS_2$} 
\label{AdS2 example}
We now give a detailed application of the results so far. Two-dimensional
Anti-de-Sitter space, $AdS_2$, appears (with an $S^2$ factor), as the
near-horizon geometry of the extremal Reissner-Nordstr\"om
black-hole, and seems to encode some of the properties of the latter
\cite{GibbonsandTownsend}. Moreover,type 0A string
theory on $AdS_2$ is conjectured  to be related to matrix models
\cite{Strom1}. The inequivalence of certain quantum field theory vacua
in Anti-de-Sitter spaces often plays a major role in both the
semi-classical study of black-hole physics and the $AdS/CFT$
correspondence.  We will see that the mathematical tools introduced so
far and the notions of change of inertial frame
and change of observers of Sec.\ref{physical
  interpretation} have interesting physical applications, especially
 to  distinguish the
global time, the Poincar\'e time and the Schwarzschild time of
$AdS_2$. Our remarks generalise quite simply to $AdS_n$. \\
The line element of (the universal cover of) $AdS_2$ in global
coordinates reads: 
\ben \label{AdS2 metric}
{\rm d}s^2= { R^2 \over \sin^2 \sigma }(-{\rm d}\tau^2 +{\rm
  d}\sigma^2)
\een
 where $0\leq \sigma \leq \pi$, $\tau \in \mathbb{R}$ and we identify
 $\tau \equiv \tau +2\pi$ to get $AdS_2$. We take the curvature radius
 to be $R=1$, and denote the metric by $h$ as usual. Its Killing
 vectors (with $\partial_{\tau}\equiv \partial /\partial \tau$ and
 $\partial_{\sigma}\equiv \partial / \partial \sigma$) are given by: 
\begin{align}
X^T&= {1 \over \sqrt{2}}\partial_{\tau}, \quad X^Y={1 \over
  \sqrt{2}}(\cos \tau \cos \sigma \partial_{\tau} - \sin \tau \sin
  \sigma \partial_{\sigma}) \nonumber \\
X^Z&=-{1 \over \sqrt{2}}(\sin \tau \cos \sigma \partial_{\tau} + \cos
  \tau \sin \sigma \partial_{\sigma}) \label{Killing vectors of AdS2}
\end{align}
and satisfy the commutation relations:
\ben \label{AdS com rel}
[X^T, X^Y]= {1 \over \sqrt{2}}X^Z, \quad [X^T, X^Z]=-{1\over \sqrt{2}}X^Y,
  \quad [X^Y, X^Z]= -{1 \over \sqrt{2}}X^T   
\een
Their flow defines an effective action on $AdS_2$ which in turn induces the
following Lie-algebra anti-homomorphism: 
\ben
\phi: A\equiv tT + yY+zZ \mapsto tX^T+yX^Y + zX^Z
\een
where $T$, $Y$ and $Z$ span an ${\frak sl}(2, \mathbb{R})$ Lie-algebra
  with commutation relations given by the opposite of (\ref{AdS com
    rel}), and $(t,y,z)$ are the corresponding coordinates of an
  element $A\in {\frak sl}(2, \mathbb{R})$. Let $(AdS_2, C_p)$ the Segal structure stemming from the
  metric (\ref{AdS2 metric}), where $\partial_{\tau}$  is
  defined to be future-directed. By
  Lemma \ref{connected 
  Lie actions}, the connected Lie group $SL(2, \mathbb{R})$  acts as causally
  preserving diffeomorphisms of $(AdS_2, C_p)$. \\
Now $SL(2, \mathbb{R})\simeq Sp(2, \mathbb{R})$ is in Table
  \ref{irred herm} and thus  admits
  bi-invariant cones.  One such cone is
  given by its Killing form $B: (A,B)\mapsto {\rm Tr}(ad_A \circ
  ad_B)$, where $ad_A(C)\equiv [A,C]$ is the adjoint representation of
  the Lie algebra on itself. Indeed, we can easily check that
  $B(\;,\;)$ defines a 
  Minkowski product on ${\frak sl}(2, \mathbb{R})$:
\ben \nonumber 
B(A, B)= -t_At_B +y_Ay_B+z_Az_B 
\een
Since the Killing form is invariant under the Adjoint action and the
  group is connected, 
\ben \label{cone in sl2R}
C_e\equiv \{ A\in {\frak sl}(2, \mathbb{R}) \, / \, t\geq 0, \;
  B(A,A)\equiv -t^2 +y^2 +z^2 \leq 0 \, \}
\een
defines a (pointed closed convex) cone invariant under $Ad_{SL(2,
  \mathbb{R})}$. Let $p\in AdS_2$ the point with coordinates
  $(\tau=0, \sigma=\pi /2)$, and let $A=tT +yY+ zZ \in C_e$. Then $X^A_p = {1 \over \sqrt{2}}(t\partial_{\tau}
  -z\partial_{\sigma})$ satisfies 
\ben \label{AdS2 causal}
h_p(X^A_p, X^A_p)={1 \over 2}(-t^2+z^2) \leq 0
\een
and is future directed. We clearly have $\mu_p\ast
(e)(C_e)=C_p$. Corollary \ref{obs indep cor} then implies  that the
Segal structure  $(AdS_2, C_p)$ is
  observer-independent. It is static of course since $X^T_p \in
  Int(C_p)$ for all $p \in AdS_2$. However, observer-independence of
  the Segal structure does not imply that all future directions are
  equivalent: it somply means that any future-directed vector can be
  extended to a global causal Killing field. Comparing (\ref{cone in sl2R}) and
  (\ref{AdS2 causal}) we see that there are some $A\in C_e$ such that
  $B(A,A)=0$   but $h_p(X^A_p, X^A_p)<0$: some 
  ``light-like''  vectors in the group can be mapped to stricly
  time-like ones on spacetime. 

This remark suggests to take a closer look at the ``light-like''
vectors in the Lie algebra. To do so,  let us compare the globlal
static Killing field 
$\partial_{\tau}$ and
the one which defines time in Poincar\'e coordinates. Let   
$x\pm z \equiv \tan((\tau\pm \sigma)/2)$, or equivalently:
\ben \nonumber
x={ \sin \tau \over \cos \tau + \cos \sigma}, \quad z={\sin \sigma \over
  \cos \tau + \cos \sigma}   
\een These coordinates only cover $AdS_2$ by patches, since they
break down when $\tau= \mp \sigma +(2k+1)\pi$ ($k \in \mathbb{Z}$). We
have $x \in \mathbb{R}$ and $z\geq 0$ or $z\leq 0$ according to the
patch. In
each of these patches, the metric (\ref{AdS2 metric}) reads ($R=1$):
\ben 
{\rm d}s^2 = {R^2 \over z^2}(-{\rm d}x^2 +{\rm d}z^2)
\een
with 
\ben \label{Poincare Killing field}
{\partial \over \partial x}= (1+ \cos \tau \cos \sigma)\partial_{\tau}
-\sin \tau \sin \sigma \partial_{\sigma}=\sqrt{2}(X^T+X^Y)
\een 
The Poincar\'e Killing field $\partial / \partial x$ is causal and has a horizon
at $z=+\infty$. In fact (\ref{Poincare Killing field}) is true in all
the coordinate patches, and $\partial / \partial x$ can be uniquely
globally extended to 
 the image of $H=
\sqrt{2}(T+Y)\in C_e$ under $\phi$. Thus it is causal everywhere,
though not static. Indeed,   
\ben \nonumber
h_p(X^H_p,X^H_p)= - { (\cos \tau + \cos \sigma)^2 \over \sin^2 \sigma
} \leq 0
\een
and vanishes on the (Killing) horizons between the patches. \\
Proposition \ref{universal algebra
  prop} says that the global field $X^T$ and the Poincar\'e field
$X^H$ can define observables which are related by a change of inertial
frame if and only if the vectors $T$ and $H$ are Adjoint related in
${\frak sl}(2, \mathbb{R})$. However, $B(T,T)<0$ whereas $B(H,H)=0$,
so $T$ and $H$ cannot be adjoint related. Thus, the global and
the Poincar\'e Killing fields {\it do not define equivalent
  observables}. Note also that  if we restrict $AdS_2$ to one
Poincar\'e patch, the vector $H$ defines a
static Killing field $X^H$, though it is in the boundary of the cone $C_e$.

Moreover, some ``space-like'' vectors in the Lie algebra can
define Killing fields which become causal in certain regions. These in
fact are related to  what is called in \cite{Strom2} the Schwarzschild
time of $AdS_2$:  
the Schwarzschild-type coordinates of $AdS_2$ are those in which the
preferred causal 
Killing direction corresponds to the prefered time direction at
spatial infinity in the near-extremal Reissner-Nordstrom black
hole --of which $AdS_2\times S^2$ is the near horizon geometry--. The time in the near-horizon geometry is thus that which is
taken to define the Boulware vacuum in the black hole space
\cite{Strom2}. In the coordinate patch $-{\pi \over 2} \pm \sigma \leq \tau \leq {\pi
  \over 2} \pm \sigma$, let  
\ben \nonumber
\tan {1 \over 2}(\tau \pm \sigma \mp {\pi \over 2}) = \mp e^{\mp q -\rho}, 
\een
so that the $AdS_2$ metric (\ref{AdS2 metric}) reads:
\ben \nonumber
{\rm d}s^2= { R^2 \over \sinh^2 \rho}(-{\rm d}q^2 + {\rm d}\rho^2)
\een
where $q\in \mathbb{R}$ is the Schwarzschild time and $\rho > 0$
represents the inverse distance to the horizon of the black hole. The horizon of $\partial /\partial q$ at $\rho
=+\infty$ corresponds to the black hole horizon. In fact, this horizon
is precisely to $AdS_2$ what the Rindler horizon is to Minkowski
space.\footnote{Rindler space appears (times $S^2$) as the
  near-horizon limit of the Schwarzschild solution, and  Rindler time
  corresponds to $t$ in (\ref{B-H metric}). \label{Rindler footnote}
  In
  Minkowski space the Rindler Killing vector $\partial /\partial t$
  is a velocity boost generator, which is also  ``spacelike'' (or
  non-compact) in ${\frak so}(1,1)$.}  
One can check that the Killing field $\partial /
\partial q$ is equal to $\sqrt{2}X^Y$ on the coordinate patch, so that
it can be 
uniquely globally extended to that Killing field. However,
$\sqrt{2}X^Y$  is not
static nor causal; its  horizon corresponds to that of $\partial / \partial
q$. Moreover, $B(\sqrt{2}Y, \sqrt{2}Y)=2 > 0$ and $Y$ is
``space-like'' in ${\frak sl}(2, \mathbb{R})$. It can neither be
mapped to $T$ nor to $H$ via an Adjoint action. The Schwarzschild
field $\partial / \partial q$ defines a third non-equivalent
time and corresponding energy observable; the three observables
cannot be related by inertial transforms.  \\      

In quantum theory, the respective Hamiltonians of the global, the
Poincar\'e and the Schwarzschild times, will
not always have the same eigenvalues; states with positive frequencies
with respect to one time variable cannot always
be mapped unitarily to states with positive frequencies with respect
to the other time variables. The horizons of $X^H$ and $X^Y$ differ in
nature, since $X^H$ 
remains causal globally, whereas $X^Y$ becomes spacelike and time-like
past-directed in certain regions. We expect the Hamiltonian of $X^H$
to be better behaved than that of $X^Y$. We will come bak to this 
in Sec.\ref{positive energy}. 
All the ``strictly  time-like'' vectors in $C_e$ can however be
mapped to $T$ (up to scale) via an Adjoint action, and similarly the
``light-like'' ones to $H$ and the ``space-like'' ones to
$Z$ (up to sign): as we know, the elements of ${\frak sl}(2,
\mathbb{R})$ fall into three
types more often called the compact ones (like $T$), the parabolic
ones (like $H$),  and the
non-compact ones (like $Z$).  There
exists inequivalent observers in $AdS_2$ with (local) times corresponding to
each of these types.

\section{Horizons in the group of motions}
\label{horizons in the group of motions}
\subsection{Definitions}
We now consider general spacetimes with Segal structures which are
not necessarily 
observer-independent or static, and develop a novel group 
theoretical description of the horizon structure of
spacetimes, which takes into account the notions of change of inertial
frame and 
change of observer of Sec.\ref{physical interpretation}. Horizons
occur in a spacetime whenever a Killing field is not static. Let $(\M,
h)$ a time-orientable (connected) Lorentzian manifold equipped
with its Segal structure and its connected symmetry group $G$ as
in Lemma \ref{connected Lie actions}.
\begin{Def} \label{horizon on M def} Let $X^A$ a Killing vector field of $(\M, h)$, and
suppose $X^A_p\in Int(C_p)$ for some $p\in \M$. The horizon of an
observer at $p$ with future $X^A_p$ is the following set:
 \ben \label{horizon on M}
 \mathcal{X}_{A}\equiv \{\, q \in \M \,/\,
h_q(X^A_q, X^A_q)=0 \,\}
 \een
 It is empty if and only if $X^A$ is a static
Killing field.
\end{Def}
(We could also define the horizon of $X^A$  for any Segal
structure as the set of points $q\in \M$ where  $X^A_q$ is in the
topological boundary of or is extreme in $C_q$). Definition \ref{horizon on M def}
includes, as particular cases, Killing horizons \cite{Carter} such as
the $r=2M$ surface   \cite{HawkingandEllis} of the Schwarzschild solution (\ref{B-H metric}),
cosmological horizons \cite{HawkingandEllis} such as those of
de-Sitter space, boundaries of
Ergo-regions such as in the Kerr solution \cite{Penroseergo},
the Rindler horizon of Minkowski space\ldots These features of the causal
structure of spacetimes are related to physical properties such as
Hawking radiation or possible energy extraction from a region of
spacetime. In
the physical applications, the choice of a Killing field to extend a given 
causal vector at a point is not arbitrary; it is in fact decisive.  In
Minkowski space $\mathbb{E}^{1,3}$ with  flat
coordinates $(t, x,y,z)$, the Killing vector fields $\partial_t$
and $\partial_t +y\partial_x-x\partial_y$ coincide at the origin,
 but only $\partial_t$ is static. The horizon
of $\partial_t +y\partial_x-x\partial_y$ is generally not thought to be
relevant. However, the example of $AdS_2$ in Sec.\ref{AdS2 example}
showed that even in a spacetime with a static or observer-independent
causal structure, horizons which at first seem to appear from a
bad choice of time coordinate, may in fact be given a physical
interpretation --a black hole horizon in this case--. 
The Rindler horizon of Minkowski space  is similar 
example (see footnote p.\pageref{Rindler footnote}). The fact that the
horizon of  any Killing field of $(\M,h)$ should be
relevant to 
an observer stationary with respect to this field, will be motivated
in quantum theory in Sec.\ref{positive energy}. \\
For homogeneous spacetimes, we can  lift $\mathcal{X}_{A}$ to a
locus in $G$:
\begin{Def} \label{horizon in G def} Let $(\M, h)$ a homogeneous Lorentzian spacetime, and
let $G$ its connected group of motions as in Lemma \ref{connected
Lie actions}.  For $X^A$ is a Killing vector field of $(\M,h)$
which is timelike future-directed at $p$, we define the horizon in
$G$ for an observer at $p\in \M$ with future $X^A_p$ to be the set
\begin{equation}
\label{horizon in G} \H_{A,p}\equiv \{\, g\in G \,/\;h_{g.p}(X^A,
X^A)=0 \,\}= {\mu_p}^{-1}(\mathcal{X}_A)
\end{equation}
\end{Def}
$\H_{A,p}$ is also defined when the spacetime is not homogeneous;
however, it can be empty when $\mathcal{X}_A$ is not. This occurs
typically in black-hole spacetimes such as (\ref{B-H metric}),
where the group action cannot move along the radial direction, so
that the $G$-orbits through points outside the black-hole do not
intersect its horizon. In these cases, the near-horizon geometries might
be homogeneous and encode some of the horizon structure. 

 $\mathcal{X}_A$, in a homogeneous spacetime, is often thought as
a surface or rather a cylinder surrounding an observer at $p\in
\M$. The region of 
spacetime inside the horizon $\mathcal{X}_A$, is that for which
particles stationary with respect to the observer with future
$X^A$  have
causal trajectories. $\H_{A,p}$ on the other hand can be thought
as a surface in $G$ ``surrounding'' the identity element $e$. 
Definition \ref{horizon in G def} enables us to  ``center'' at the
same point $e\in G$ all the horizons in $\M$ for different
observers  with different future directions. As we shall see, it
turns out that these different horizons in $G$ can be elegantly
compared. To do this, we first lift the spacetime metric $h$ to a degenerate
metric on $G$.

\subsection{Degenerate metrics on $G$}
Given $(\M,h)$, the metric tensor field $h$ pulls back to $G$ using for fixed
$p\in \M$ the map $\mu_p:G\to \M$. Define $\tilde{h}^p
\equiv {\mu_p}^{\ast}(h)$ for fixed $p$ so that
 \begin{align*}
\forall \, X_g,Y_g \in T_gG, \quad \widetilde{h^p}_g(X_g, Y_g)
&\equiv
[{\mu_p}^{\ast}(h)](X_g, Y_g) \\
&\equiv h_{g.p}(\mu_p \ast(g) X_g, \mu_p \ast(g) Y_g)
\end{align*}
The Killing vector fields correspond on each $G$-orbit in $\M$ to
the images by $\mu_p$ of the {\it right}-invariant fields
$\tilde{A}_g=R_g \ast (e)A$ on $G$. In contrast, the metric
$\widetilde{h^p}$ on $G$ is {\it left}-invariant. Indeed, for $A,B
\in {\frak g}$, using $\nu_g \circ \mu_p=\mu_p \circ L_g$ and
equation (\ref{killingisometry}), we get
\begin{align*}
\widetilde{h^p}_g (L_g \ast (e) A, L_g \ast (e) B) &\equiv
h_{g.p}(\mu_p \ast(g)L_g \ast (e) A , \mu_p
\ast(g) L_g \ast (e) B) \\
 & \equiv h_{g.p}(\nu_g \ast (p) \mu_p \ast (e) A, \nu_g \ast (p) \mu_p \ast (e)
 B) \\
 &=h_{g.p}(\nu_g \ast (p) X^A_p, \nu_g \ast (p) X^B_p )=h_p (X^A,
 X^B)\\
  & \equiv \widetilde{h^p}_e(A,B)
 \end{align*}
 and hence ${L_g}^{\ast}\widetilde{h^p}=\widetilde{h^p}$.
Thus $\widetilde{h^p}$ is simply determined by its behavior on
${\frak g}\simeq T_e G$. For convenience we simply denote by
$<\;,\;>_p$ the corresponding inner product on ${\frak g}$ so that
\begin{equation}
\label{pull-back} \forall \, A,B\in {\frak g}, \quad <A,B>_p
\equiv h_p(X^A, X^B).
\end{equation}
Since $h$ is a non-degenerate Lorentzian metric, the kernel of
$<\;,\;>_p$ defines an (null) isotropic subspace of $h$ in $T_p\M$
which has dimension 1 at most.  The map $\mu_p\ast(e)$ has kernel
${\frak h}_p$ the Lie algebra  of the stabiliser subgroup $H_p$,
thus the Kernel of $<\;,\;>_p$ has dimension $(dim\; {\frak
h}_p)+1$ at most. When $\M$ is homogeneous, the $\mu_p\ast (e)$
are surjective and $<\;,\;>_p$ has kernel ${\frak h}_p$ precisely, so
that if $G$ acts simply transitively (i.e. ${\frak h}_p=\{0\}$),
$<\;,\;>_p$ defines a left-invariant Lorentzian metric on $G$. 

\subsection{Horizons in $G$ and their properties}
Using (\ref{killingisometry}) and (\ref{adjoint c o f}), the inner
products $<\;,\;>_p$ on ${\frak g}$, for different $p\in \M$,
satisfy:
\begin{equation}
\label{adjoint pull-back} \forall \, A,B \in {\frak g}, \quad
<A,B>_{g.p}= h_{g.p}(X^A, X^B)=<Ad_{g^{-1}}A, Ad_{g^{-1}}B>_p
\end{equation}
This equation gives the metric product of Killing vector fields of $(\M,
h)$ as one moves along the $G$-orbit of a point $p\in \M$. The
non-commutativity in $G$, encoded in the Adjoint action, becomes essential. 
Equation (\ref{horizon in G}) then reads:
\begin{equation}
\label{horizon in G 2} \H_{A,p}\equiv \{ g\in G
\,/\;<Ad_{g^{-1}}A, Ad_{g^{-1}}A>_p=0 \,\},
\end{equation}
and $\H_{A,p}$ is completely defined group theoretically.
As a consequence, the
identity component of the
centre of $G$, as that of $H_p$, is ``inside'' the group horizon
$\H_{A,p}$. The same is true, by continuity
of the map $g\mapsto <Ad_{g^{-1}}A, Ad_{g^{-1}}B>_p$, of all elements of a
sufficiently small neighborhood of the identity element $e\in G$. \\
More importantly, equation (\ref{horizon in G 2}) underlines the
physical implications of the existence of (possibly degenerate)
{\it bi-invariant} Lorenztian metrics on Lie groups. Indeed, these can
then be restricted to a homogeneous space $\M\simeq G/H$ on which they are
non-degenerate. The corresponding Segal structure is
then trivially observer-independent, static, and all the Killing fields
which are causal at one point have no horizons.     
This is the case for Lie groups admitting bi-invariant Lorentzian
metrics, so for the maximally supersymmetric backgrounds of
chiral supergravity in six dimensions \cite{CFS}. In these
backgrounds, the Killing fields stemming from the Lie group
description have no horizons, but this does not preclude extra
Killing fields from having horizons.  
 We can now easily express
how horizons behave under changes of inertial frames.
\begin{Prop}
\label{killing hor prop} Let $\H_{A,p}$ the horizon in $G$ of an
oberserver at $p\in\M$ with future $X^A_p$. Then for all $g, k \in
G$ we have
 $\H_{Ad_gA,k.p}=g\H_{A,p}k^{-1}$,
   which implies: \\
(i) $\H_{Ad_gA,g.p}=g\H_{A,p}g^{-1}$, so that changes of inertial
frames in $\M$ induce group conjugations on the horizons in $G$,\\
(ii) $\H_{A,k.p}= \H_{A,p}k^{-1} $, so that the horizons of
observers with the same future are related by group translation.
\end{Prop}
{\it Proof:} Using equation (\ref{adjoint pull-back}) and
(\ref{horizon in G 2}), we have
\begin{align*}
h\in \H_{Ad_gA,k.p}& \Leftrightarrow < Ad_{h^{-1}}Ad_gA,
Ad_{h^{-1}}Ad_g A>_{k.p}= 0\\
 & \Leftrightarrow < Ad_{k^{-1} h^{-1}g }A,
Ad_{k^{-1}h^{-1}g} A>_{p} =0 \\
 & \Leftrightarrow g^{-1}h k \in \H_{A,p} \Leftrightarrow h\in
 g\H_{A,p}k^{-1} \qquad \Box
\end{align*}
With the inverse exponential, we can further lift the horizons to
${\frak g}$, and show that conjugation in {\it (i)} becomes the Adjoint
action.  
 Property {\it (ii)} for
$h\in H_p$ implies that one can map $\H_{A,p}$ to $\M\simeq G/H_p$,
 and get a bijection between $\H_{A,p} / H_p$ and $\mathcal{X}_A$, as
 expected. {\it (i)} says that any two inertially related
observers in $\M$  have  conjugate horizons in $G$ (so in
bijection).
\footnote{Analogously, all of our line-of-sight 
 horizons on earth define conjugate loci in $SO(3)$ !}  
At a given spacetime point $p\in \M$, the inequivalent future directions
classify as the projective orbits of ${\mu_p}^{-1}(Int(C_p))\equiv \{A\in {\frak g}/
X^A_p \in Int(C_p) \}$ under the action of $Ad_{H_p}$. For any $A\in
\mu_p^{-1}(Int(C_p))$ and $h_1,h_2\in H_p$, we have
\ben \nonumber
\H_{Ad_{h_1}A,h_2.p}=h_1\H_{A,p}{h_2}^{-1}=h_1\H_{A,p}
\een
Thus the set of inequivalent horizons 
for observers at $p\in \M$ is in one-to-one correspondence with the
set of orbits of $\{\H_{A,p} / A\in {\mu_p}^{-1}(Int(C_p))
\}$ under left multiplication by $H_p$. 
For instance, since  the Poincar\'e and
the Schwarzschild Killing fields are not even Adjoint related in
${\frak sl}(2,\mathbb{R})$, their respective horizons in $AdS_2$ are
neither equivalent for observers at the same point, nor equivalent for
inertially related observers. \\
 Proposition \ref{killing hor
prop} should shead a light on the nature of physical properties of
horizons in a spacetime. These features should be invariant under changes of
inertial frames, and transform covariantly under changes of observers. 
When lifted  to the symmetry group, they are expected to transform
accordingly under group conjugation or translation. For example, the
expressions  
for the de-Sitter temperature \cite{GibbonsandHawking} and the Unruh
temperature \cite{Unruh}, when
transcribed in this group theoretical formalism, must be invariant
under group conjugation.  \\ 
Equation (\ref{horizon in G 2}) shows how horizons can be thought of
as arising merely from the composition law of the symmetry group of
spacetime: the metric at one spacetime point $p\in \M$ together with
the Adjoint action of the group, suffice to determine completely the
horizons of all 
observers in $\M$. This gives an elegant description of the
horizon
structure of de-Sitter space or that of   
the Poincar\'e Killing fields of Anti-de-Sitter space. Moreover,
this definition might be
helpful to study the topology of horizons.  

\section{Positive energy and bi-invariant cones} \label{positive
  energy}
We now look at the relations between properties of infinitesimal
causal structures on a manifold and the possibility to define a
positive energy for states of a physical theory with spacetime
interpretation. We say that a theory has positive energy
whenever there exists an energy observable whose associated functional
is one-side bounded. We shall see that for
quantum theories, 
the existence of particular invariant cones in the symmetry group
yields  a
necessary and sufficient condition.   
\subsection{Causal Killing fields and observer-independent classical
  observables} 
\label{causal killing fields and energy} 
We consider a time-orientable
Lorentzian spacetime $(\M, h)$, together with its group of Killing
symmetries as in Lemma \ref{connected Lie actions}.
We already mentioned that
Killing symmetries of $(\M, h)$ yield conserved quantities along
the geodesics: given a geodesic $\sigma \mapsto
(x_{\mu})(\sigma)$ and a Killing vector field $K^{\mu}$, the
quantity $K^{\mu}Dx_{\mu}(\sigma)/D\sigma$ does not depend on
$\sigma$. It is usually called energy, momentum, or angular
momentum, according to whether $K^{\mu}$ generates a time
translation, a space translation or a rotation in $\M$.  Letting $K^{\mu}\equiv X^A$, the functional
 \ben \label{energy functional}
V_p\mapsto h_p(X^A_p, V_p)
 \een
is positive on $C_p$ for all $p\in M$ if and only if $X^A$ is causal.
It can be composed with $\Psi_p$ in  diagram (\ref{general theory comm
diagram}) to define an  observable-functional on the states of a general
theory. However, any causal vector field $T$ on
$(\M,h)$ similarly defines a negative  
functional. Why should these be discarded ? In fact in general relativity
there is no reason to priviledge even a
particular {\it local} future from another one. The major discrepancy between classical and quantum
theories stems from the difference between local and global
descriptions. A theory of causal curves does not need a global
future or global observables: time evolution has a sense locally. As
soon as as one goes to a global picture, one 
tends to replace the notion of localized observer by that of 
reference frame or ``family of observers'',  and hence think that the
time parameter in the Schr\"odinger-type equation should correspond to
a  global future in spacetime. This is misleading. The time parameter
of the unitary evolution need not be a global future on spacetime in
the sense of 
causal vector fields; it is attached to a particular observer. The
specificity of Killing fields, as we shall
show in the next section, is precisely that they all allow for a {\it Schr\"odinger-type
evolution} in the Hilbert space. In that sense, whether they are
(globally) causal
or not, they are preferred mathematical tools to define time evolution in
quantum theory, for particular 
observers. \\
Now the fact fact that the Killing fields should be causal is another
issue. Well-defined time evolution does not imply positive energy. If
one accepts that,
in our universe,
future-directedness can be an observer-dependent concept --by which we
mean that 
the Segal structure of our spacetime need not be
observer-independent--, then the  
unboundedness  of
the energy simply comes with it. Such an assumption is not absurd. It
is evident that given a
particle with velocity $V_p$ at $p\in \M$, 
two observers at $p\in \M$ boosted with respect to each other measure different
values for (\ref{energy functional}). Now whatever the spacetime,
their values have the same
sign of course. For observers at different points, these values might
have different signs: this just means that mathematically we are not
able to define a common Killing future direction, and hence define the
{\it same} unitary evolution in the Schr\"odinger picture for both
observers. However, in the real world, an
observer at $q\neq p$ cannot measure anything at $p$, and thus such
measurements have no meaning. If we think that an observer
in de-Sitter  space cannot measure a
state at his antipodal point, there is no
paradox in having unbounded energy functionals. It becomes a
problem of interpretation. \\

 Lemma \ref{causal killing field lem} gives the necessary and
sufficient conditions for a spacetime $(\M, h)$ to admit a bounded
energy functional $V_p\mapsto h_p(X^A_p, V_p)$, and the existence of
an Adjoint-invariant cone in the symmetry group  is required. \\
Furthermore, equation (\ref{killingisometry}) implies  the values of all
observables (\ref{energy functional}) on a given state represented
by $V_p\in C_p$, remain the same
 under changes of inertial frames: both the
state and the observables are mapped with $\nu_g\ast(p)$ to
physically equivalent ones. Of course this is not true for changes
of frames induced by conformal transformations of $(\M, h)$.
 On
the other hand requiring that these functionals be invariant under changes of observers highly
restrains the possibilities. By this we mean that the outcome on a
given state represented  by
$V_p\in C_p$ does not vary under the changes of observers of Sec.\ref{changes of observers}: the Killing field $X^A$ must be
such that for all $p\in \M$ and $V_p\in T_p\M$ fixed, we have
  \begin{align}
   \forall g\in G, \quad  h_{g.p}( X^A_{g.p}, \nu_g \ast (p)V_p) &= h_p(X^{Ad_{g^{-1}}A}_p,
 V_p)  \nonumber \\
 \label{obs indep functional} & = h_p(X^{A}_p,
 V_p).
  \end{align}
 \begin{Lem} \label{obs indep functional lem} Let $G$ a maximal (connected) Lie group act effectively on
 an $n$-dimensional spacetime $(\M, h)$ as its Killing motions. Then
 the observable $V_p\mapsto h_p(X^A_p, V_p)$ is invariant under changes of observers if and only $A\in
 {\frak g}$ is invariant under the Adjoint action of $G$, or
 equivalently $A$ is in the center of ${\frak g}$. This implies that
 for all $p\in \M$ such that $X^A_p\neq 0$, $H_p\subset SO(n-1)\subset SO(1,
 n-1)^+$, so that there are no ``boost symmetries'' at this point.
 \end{Lem}
 {\it Proof:} Condition (\ref{obs indep functional}) is satisfied
 trivially when $A\in {\frak g}$ is $Ad_G$-invariant.
 Since  $h_p$ is non degenerate,
 \ben \{ X \in T_p\M \,/\, \forall V_p\in T_p\M, \, h_p(X^{A}_p+X,V_p)
 =h_p(X^{A}_p,
 V_p) \, \}=\{0\}. \nonumber
  \een
  As a consequence, for all $p\in \M$ and
 $g\in G$, $Ad_{g^{-1}}A-A\in {\rm Ker}(\mu_p\ast(e))$. But
 $\bigcap_{p\in \M}{\rm Ker}(\mu_p\ast(e))=\{0\}$ for an effective
 action, so that $A\in {\frak g}$ is $Ad_G$-invariant.  \\
 Let $X^A$ such a Killing field,  and
 suppose $X^A_p\neq 0$ and $X^A_p\notin C_p$. Let $V_p\in Int(C_p)$, so
 that there exists $\lambda>0$ such that $V_p\pm \lambda X^A_p\in
 Int(C_p)$. If $H_p$, or rather the linear isotropy group $\{ \nu_h\ast (p)\,/\, h\in
 H_p\}$, contains a boost symmetry, then there exists $h\in H_p$
 such that $\nu_h\ast (p) V_p$ is arbitrarily close to the
 boundary of $C_p$. (The boosts of $SO(1,n-1)^+$ take points ``up'' the hyperbolae in $C_p$,
 thus arbitrarily close to the boundary.) Then $h\in H_p$ can be
 chosen such that either of
  $\nu_h\ast(p)(V_p\pm \lambda X^A_p)=\nu_h\ast (p) V_p\pm \lambda
  X^A_p$ is not in $C_p$, which is a contradiction. If $X^A_p\in C_p$, take
  $-X^A_p$.
 $\Box$ \\
 For example, as we noticed earlier,
 $\partial_t$ in the Schwarzschild metric (\ref{B-H metric}) --or
 rather its antecedent by $\phi$-- is invariant under the Adjoint action of the symmetry
 group  $\mathbb{R}\times SO(3)$. The corresponding energy functional,
 though it is  not bounded on the whole black-hole spacetime, is
 invariant under changes of observers (and changes of inertial frames of course) which relate any two different
 points with the same radial coordinate. This is not true of
 $\partial_{\psi}$, so that angular momentum is not observer
 independent. \\
 Of course this notion of invariance under changes of observers is itself
 invariant under changes of inertial frames.  Spacetimes with
 simple groups of motions $G$ (with $dim\,G\geq 2$) cannot
 admit such Killing fields $X^A$, because $A$ would span an ideal
 in ${\frak g}$. This excludes for example Anti-de-Sitter and de-Sitter spaces, all Bianchi models with simple
 maximal symmetry groups\ldots The ``no-boost'' condition also excludes many
 spacetimes, the simplest being  Minkowski space
 $\mathbb{E}^{1,n-1}=E(1,n-1)/SO(1,n-1)$. Of course for physical
 relevance, the group of motions to consider is that induced by the
 symmetries of  particular theory. \\
To go from the
classical notion of positive energy considered in this section, to that defined by
 one-side bounded
quantum  operators, the structure of the universal algebra of the
 group of motions will be useful. 

\subsection{The universal algebra and the Dirac procedure}
\label{Dirac procedure section}
Remember that the universal algebra  of ${\frak g}$, $\U({\frak
g})$,  is the extension of ${\frak g}$ into 
an associative algebra such that the commutator of two elements of
${\frak g}$ be given by their Lie bracket. Proposition
\ref{universal algebra prop} establishes a correspondence between
$\U({\frak g})$ and particular spacetime observables, which generalise
those associated to Killing vector fields. These can be thought of as
a set of differential operators on the spacetime $\M$, and constitute
an infinite-dimensional algebra which is fundamentally
non-commutative. \\
The Dirac procedure is usually defined between the Poisson algebra of
real valued functions  on phase space, and an associative algebra of
operators acting on a Hilbert space. It suffers from ordering
problems. In our approach, we will define a  Dirac map from $\U({\frak
  g})$ to a set of operators. There are no ordering problems at this
point since
$\U({\frak g})$ is already a non-commutative associative algebra. 

\subsubsection{Quantum observables and changes of inertial frames}

 Let ${\frak
  g}$ the real Lie 
algebra of a group of symmetries $G$ acting almost-effectively on a
space $\M$. $\hbar$ is Planck's constant divided by $2\pi$, and $i$ is
$\sqrt{-1}$.    
\begin{Def} \label{Dirac procedure def} Let $\pi \ast: {\frak g}\to {\rm AHerm}(\mathcal{F}) $ a
  representation of ${\frak g}$ into a set of
  anti-self-adjoint operators 
  of a Hilbert space $\mathcal{F}$. The Dirac map 
  $\widehat{\phantom{U}}$ is defined as: 
\begin{align*}
\widehat{\phantom{U}}: {\frak g} &\longrightarrow {\rm Herm}(\F) \\
                             A &\longmapsto \widehat{A}\equiv
  i\hbar \pi\ast(A)
\end{align*}
so that for all $A,B\in {\frak g}$ we have:
 \ben \label{Dirac map com rel}
 \widehat{\phantom{U}}({[A, B]})= {1 \over i\hbar}[\widehat{A},
\widehat{B}]\equiv {1 \over i\hbar}(\widehat{A}\widehat{B}
 -\widehat{B}\widehat{A}). 
 \een
\end{Def}  
According to the point of view, the Dirac map is induced by a unitary
 representation 
 $\pi$ of $G$, or it defines through exponentation a representation of
  $G=\exp {\frak
 g}$. For $A\in {\frak g}$ and $v\in \F$, we have:
\ben  \label{derived repr}
\pi\ast(A)v\equiv {{\d} \over {\d}t} \pi(\exp tA)v \arrowvert_{t=0}
\een
and 
\ben \nonumber
\pi({\exp A})v=\exp(\pi\ast(A))v
\een
There are issues relating to convergence and definition of these maps, and generally
 speaking, the derived representation (\ref{derived repr}) is only
 defined on a dense  subset of
  $\F$, the pre-Hilbert space of vectors $v\in \F$ such that
 $g\in G\mapsto \pi(g)v$ is smooth \cite[p.388]{Neeb}. We shall not
 dwell on these
 problems though. Physically, the  representation space $\F$ corresponds to the
 set of pure states of
 the theory.\footnote{As mentioned before, in order to have real
 linear expectation value functionals, we should be considering
 density matrices over $\F$. However, for 
 stationary observers, evolution will be unitary, so that pure states
 remain pure.} There is
 a unique way \cite[p.90]{Helgason} of extending the representation
 $\pi\ast$ of ${\frak g}$ to a 
 representation of $\U({\frak g})$ (and a representation of the latter
 determines $\pi\ast$ completely). One then
   has, for all ${A, B}\in \U({\frak g})$,
 $\pi\ast(AB)=\pi\ast(A)\pi\ast(B)$, and 
\ben \label{universal multiplication}
\widehat{AB}={1 \over i\hbar }\widehat{A}\widehat{B}.
\een
The composition law on the observables that of $\U({\frak g})$ up to
an  important factor. We denote by  $\widehat{A}\mapsto 
 {\widehat{A}}^{\dagger}$ the adjoint operation on suitable operators,
 which is defined with respect
 to the fixed scalar product on $\F$. Using (\ref{universal
 multiplication}), for $A\in {\frak g}$ and $A^n\in
 \U({\frak g})$, we have:
\ben \nonumber
\big[\widehat{A^n}\big]^{\dagger}=  (-1)^{n+1}\widehat{A^n}.
\een
Now if for physical reasons one should be allowed to
compose  the quantum (Killing) observables $\widehat{A}\in \widehat{\frak g}$  
with themselves, then we need to
extend $\pi \ast$ by 
 linearity to the complexified Lie algebra ${\frak g}_{\mathbb{C}}\equiv
 {\frak g}\oplus i{\frak g}$: indeed, as an immediate consequence of
 (\ref{universal multiplication}),
\ben \nonumber
\big(\widehat{A}\big)^n=\widehat{\phantom{U}}\big((i\hbar)^{n-1}A^n
\big),
\een
which is self-adjoint for all $n$, contrary to $\widehat{A^n}$.
Complex conjugation in ${\frak g}_{\mathbb{C}}$ preserves
 the Lie bracket, whereas the `dagger' operation $\widehat{A}\mapsto
 {\widehat{A}}^{\dagger}$ changes the sign of the
 commutator. Compatibility is recovered by defining the `$^{\ast}$'
 conjugation on ${\frak g}_{\mathbb{C}}$ as $A+iC\mapsto
 [A+iC]^{\ast}\equiv -A+iC$, which  extends to $\U({\frak
 g}_{\mathbb{C}})$ in a way that $(AB)^{\ast}=B^{\ast}A^{\ast}$ for
 all 
 $A,B\in \U({\frak g}_{\mathbb{C}})$. Then $\pi\ast$ defines a hermitian
 representation 
  of the algebra $\U({\frak 
  g}_{\mathbb{C}})$
  \cite[p.30]{Neeb}, and for all $A\in
  \U({\frak g}_{\mathbb{C}})$, 
\ben \label{hermitian rep}
\widehat{\phantom{U}}({A}^{\ast})=-\big[\widehat{A}\big]^{\dagger}
\een
Physically we are only interested in self-adjoint operators,
i.e. those which yield real valued functionals, but the unique
extension of $\pi\ast$ 
 to $\U({\frak g})$
 does not generically yield such operators. In the general case,
  (\ref{hermitian rep}) says 
 it suffices to take elements $A\in \U({\frak g}_{\mathbb{C}})$ such that
$A^{\ast}=-A$. Call ${\rm Herm}(\U({\frak g}_{\mathbb{C}}))$ the set
 of such elements.  As a real vector space, ${\rm Herm}(\U({\frak
   g}_{\mathbb{C}}))$  is in fact isomorphic to
 $\U({\frak g})$. Indeed, by using the Poincar\'e-Birkhoff-Witt
 theorem \cite[Theorem 2.1.11]{Dixmier}, one gets:
\ben \nonumber
\U({\frak g}_{\mathbb{C}})=\U({\frak g})\oplus i\U({\frak g})
\een
so that an element $A=X+iY$ is such that
$A^{\ast}=-A$ if and only if $X=-X^{\ast}$ and $Y^{\ast}=Y$. Then the
map
\begin{align} \nonumber 
{\rm Herm}(\U({\frak g}_{\mathbb{C}})) &\longrightarrow \U({\frak g}) \\
      X+iY  &\longmapsto X+Y \label{anti-self-conjugate isomorphism}
\end{align}
 is an isomorphism since the elements of $\U({\frak g})$ decompose
 uniquely into a sum of self-conjugate and an anti-self-conjugate elements. 
Thus although we have complexified ${\frak g}$ to ${\frak
 g}_{\mathbb{C}}$, we still essentially have the same classical
 observables. Basically in $\U({\frak g})$ 
 these are polynomials of elements of ${\frak
  g}$ with symmetric coefficients of odd rank and anti-symmetric
coefficients of even rank, like $A+BC-CB$ 
for example, while in $i\U({\frak g})$ the symmetricity
 of the coefficients according to rank is opposite,  like in   
$i\hbar A^2$. \\   
Note that if $\pi$ were not required to be a unitary representation of
$G$, since 
  $\U({\frak g})$ 
 is itself isomorphic 
 to the algebra of left-invariant differential operators on $G$
 \cite[p.98]{Helgason}, one could take any suitable space of
 functions on $G$ as a set of physical states for our physical
 theory.  \\

We now consider changes of inertial frames in quantum theory. A
conceptual drawback of quantization procedures is that by definition
one
goes from classical to quantum theory, and hence for instance one has
to assume 
 that the quantum changes of frames stem from the classical
 ones. This is precisely the opposite to what was done in our
 discussion in Sec.\ref{changes of inertial frames}. Given this however, the
 principle of relativity implies that a quantum observable
 which
vanishes identically in an inertial frame should vanish identically in
 all inertial 
frames. In fact, we will see that this is automatically satisfied, and
show that the changes of
inertial frames as established in
Proposition \ref{universal algebra prop} for classical observables, induce  
the expected transforms on the quantum observables. We call ${\rm
   End}(\F)\equiv {\rm Herm}(\F)\oplus {\rm AHerm}(\F)$ the algebra of
 operators on $\F$ which admit an adjoint. 
\begin{Prop} \label{quantum univ prop} Let $\M$ a spacetime acted upon
  by a group of symmetries 
 $G$ with Lie algebra ${\frak g}$, and let $\widehat{\phantom{U}}$ 
a Dirac map  of a theory with symmetry $G$,  
 $\widehat{\phantom{U}}:{\frak g}\to {\rm Herm}(\F)$. Then
 $\widehat{\phantom{U}}$ extends uniquely to $\U({\frak
 g}_{\mathbb{C}})$, and for all $g\in
 G$, we have the following commutative diagram: 
  \ben \label{quantum com diag}
\begin{CD}
\U({\frak g}_{\mathbb{C}}) @ > {Ad_g } >> \U({\frak g}_{\mathbb{C}})   \\
@V{\widehat{\phantom{U}}}VV@VV{\widehat{\phantom{U}}}V   \\
 {{\rm End}(\F)}  @> {\widehat{Ad}_g} >>
 {{\rm End}(\F)}    
\end{CD}
\een
Specifically,  for all $A \in \U({\frak g}_{\mathbb{C}})$, $B \in {\frak
  g}$ and $b\in \mathbb{R}$, we have:  
\ben \label{quantum c.o.f.}
\widehat{Ad}_{\exp bB}\widehat{A} = Ad_{\exp {-ib \over \hbar}\widehat{B}} \widehat{A} =  e^{{- ib \over
 \hbar}\widehat{B}} 
 \widehat{A} \, e^{{ib \over \hbar}\widehat{B}} 
\een
The Adjoint action of the unitary operator $e^{{- ib \over
 \hbar}\widehat{B}}$ on a quantum observable, represents the effect of
a spacetime change of  inertial frame $\nu_{\exp bB}$  of type $B$ and
parameter $b\in \mathbb{R}$. Two quantum observables for different
observers 
can be thought of as physically equivalent if and only if they are on the same
$Ad_{\pi(G)}$-(projective)-orbit. 
\end{Prop}  
{\it Proof}: The fact that $\widehat{\U({\frak
    g}_{\mathbb{C}})}\subset {\rm End}(\F)$ follows from 
    (\ref{universal multiplication}) together with  $\widehat{{\frak g}}\subset
    {\rm Herm}(\F)$. Also from (\ref{universal
  multiplication}), we see that ${\rm
  Ker}(\widehat{\phantom{U}})$ is a two-sided ideal of $\U({\frak
  g}_{\mathbb{C}})$, and as such it is stable under $Ad_G$
\cite[Prop.2.4.17]{Dixmier}. 
The 
    commutative diagram simply follows. \\ 
For $A, B\in {\frak g}$
and $b\in \mathbb{R}$, we have \cite[p.118]{Helgason}:
 \ben \label{Ad and ad}
 Ad_{\exp bB}A= e^{ad_{bB}}A\equiv A+ b[B,A]+ {b^2 \over 2}[B,[B,A]]+
 \ldots,
 \een
This remains true for $A\in {\frak g}_{\mathbb{C}}$ and also for $A\in
    \U({\frak g}_{\mathbb{C}})$ (the extension is unique),   
and implies, using (\ref{Dirac map com rel}):
  \begin{align}
 \widehat{Ad}_{\exp bB}\widehat{A} &\equiv  \widehat{\phantom{U}}(Ad_{\exp bB}A) \nonumber \\
& = \widehat{A}+ {b \over i\hbar}[
\widehat{B},\widehat{A}]- {b^2 \over \hbar^2}[\widehat{B},[
\widehat{B}, \widehat{A}]]+ \ldots \nonumber \\
  & \equiv e^{ {-ib \over \hbar}ad_{\widehat{B}}}\widehat{A}=  {Ad}_{\exp
   {-ib \over \hbar }\widehat{B}}\widehat{A}. 
\label{Ad and ad quantum}
 \end{align}
Since the composition of linear operators is bilinear, (as for matrix
 multiplication),    
 this expression can be simplified to the more common form:
\ben \nonumber 
 Ad_{\exp {-ib \over \hbar}\widehat{B}} \widehat{A} =  e^{{- ib \over
 \hbar}\widehat{B}} 
 \widehat{A} \, e^{{ib \over \hbar}\widehat{B}} 
\een
The exponential of operators is defined as the usual series:
\ben  \nonumber 
e^{{-ib \over \hbar}\widehat{B}} \equiv \sum_{k=0}^{k=\infty} {{1
 \over 
k!}\Big({-ib \over \hbar}\widehat{B} \Big)^k }
\een
 It is well-defined and corresponds indeed to the Lie group
 exponential in the group of unitary 
 operators $\pi(G)$: we have $e^{{-ib \over \hbar}\widehat{B}}=
 \exp \big(b\pi\ast(B)\big)=\pi(\exp bB)$.\\
The set of self-adjoint operators  $\widehat{\U({\frak
 g}_{\mathbb{C}})}\cap {\rm Herm}(\F)$ is stable under the Adjoint
 action of $\pi(G)$.  
 Any two observables are said to be equivalent if they can be related up
 to scale by a change of inertial frame. $\Box $ \\
Proposition \ref{quantum univ prop} is the quantum analogue of Proposition
 \ref{universal algebra prop}, but whereas $\U({\frak g})$ is
 isomorphic to the algebra of spacetime (Killing) differential operators
 $\U(\phi({\frak g}))$, 
 $\pi\ast$ does not necesseraly induces a faithful representation
 $\U({\frak g})$. In a quantum theory, each spacetime symmetry
 $B\in {\frak g}$  defines 
a 
one-parameter group of unitary operators on
 $\mathcal{F}$. The elements of $i{\frak g}\subset {\frak
 g}_{\mathbb{C}}$ do not define observables, nor do they correspond to
 changes of inertial frames, since  the representation
 of the full group $G_{\mathbb{C}}=\exp {\frak g}_{\mathbb{C}}$ is not
 unitary. 

We saw in the previous section that the
 expectation values of the Killing functionals are invariant under inertial
 transforms. The situation here is similar: we
 postulate that any 
 physical state  ${\ket  u}\in \F$ for an observer
 at $p\in\M$ in a given reference frame, is taken
 under a spacetime inertial 
 transform $\nu_{\exp(bB)}$ to the
 physical state:
\ben \label{quantum vector c.o.f.}
  {\ket v} =e^{{- ib \over  \hbar}\widehat{B}}{\ket
 u},
\een 
which implies that the expectation values of the
 observables in a given state are
 invariant under inertial transform:
\ben \label{quantum invariance}
 {\bra v} Ad_{\exp {-ib \over \hbar}\widehat{B}} \widehat{A} {\ket v}= 
{\bra u}\widehat{A}{\ket u}
\een
In the classical case, the effects of changes of frames for
 both the velocity  vectors and the Killing
 observables follow directly
 from the definition of the spacetime diffeomorphisms $\nu_g$. In the
 quantum setting we
 were able to show 
 the effects of  $\nu_g$ on the observables, but need to
 assume (\ref{quantum vector c.o.f.}) to deduce the invariance law
 (\ref{quantum invariance}).\\

We now use the discussion in Sec.\ref{changes of observers} to define
changes of observers between equivalent observers on the same $G$-orbit.  
Let an observer at $p\in \M$ ``measure'' the expectation value of an
observable $\widehat{A}$  on a state $\ket{u}\in\F$, ${\bra
  u}\widehat{A}{\ket u}$, and let an observer at ${\exp bB}.p\in \M$
measure the same state with the same observable. From Proposition
\ref{universal algebra prop}, an observable $\widehat{A}$ with $A\in
\U({\frak g}_{\mathbb{C}})$ already
corresponds to a globally defined (complex) observable on $\M$, so that we do
not need to ``extend it'' to ${\exp bB}.p$. In the same way that we
had to assume the changes of coordinates on the velocity states from
$V_p\in C_p$ to 
$\nu_g\ast(p)V_p\in C_{g.p}$ (covariance rules), we postulate that the state
$\ket{u}\in\F$ in the reference frame at $p$  corresponds to the state
$e^{{-ib \over \hbar}\widehat{B}}\ket{u}$ in the reference frame at
$\exp bB.p$.  
Thus the equivalent observer at ${\exp bB}.p$ measures the expectation
value:
\ben
\bra{e^{{-ib \over \hbar}\widehat{B}}u}\widehat{A}\ket{e^{{-ib \over \hbar}\widehat{B}}u}=\bra{u} e^{{ ib \over \hbar}\widehat{B}} 
 \widehat{A} \, e^{{-ib \over
     \hbar}\widehat{B}}\ket{u}.
\een
Clearly if $\ket{u}$ did not undergo a ``coordinate
change'', changes of observers would have no effect, which is absurd. 
Equivalently, if one thinks the state $\ket{u}$ should be fixed in $\F$ but the
observables undergo a change of coordinates, then going from
${\exp bB}.p$
back to $p$ in $\M$ send  the value of the observable $\widehat{A}$ at
  ${\exp bB}.p$ to $e^{{+ib \over
    \hbar}\widehat{B}}   \widehat{A} \, e^{{-ib \over
     \hbar}\widehat{B}}$ at $p$. This is analogous to $X^A_{g.p}$ being mapped
to $X^{Ad_{g^{-1}}A}_p$. We assume that these formulae remain valid for
any observable in ${\rm Herm}(\F)$, and similarly for the formulae on
changes of inertial frames.

\subsubsection{Time evolution and bi-invariant cones}
\label{time evolution in quantum}
We now consider time
 evolution and observer-dependence in this formalism. The spacetime
 $\M$ is fixed, and it has a Segal structure $(\M,C_p)$. Relativistic
 invariance will be guaranteed throughout by the group structure.    
The quantum observables constructed through the Dirac map correspond
 to classical observables defined on all of $\M$. It is essential
 however to see what these mean at a particular
 spacetime point, in order to interpret the quantum theory for an
 observer at this point.
 The elements  of ${\frak g}$ for example 
classically  correspond to vector fields on $\M$, and their quantum
 analogues should be interpreted, {\it for a chosen observer} at $p\in \M$,
 according 
 to whether they generate time-translations, rotations, etc, {\it at
 this particular spacetime point}. Although the Hilbert space $\F$ is
 not endowed  with a
 causal structure of itself,  we will show that choosing a spacetime observer 
 amounts to doing so, since the unitary operators $e^{-{i \over
 \hbar}b\widehat{B}}$ generate future displacements 
 for an observer at $p\in\M$ only if  $X^B_p\in C_p$.  \\
Time evolution in our setting is best defined for stationary observers
 in $\M$: it merely
 corresponds to the unitary action of a particular symmetry. Consider an 
 observer in  $\M$ with future $X_p\in Int(C_p)$ at the event $p\in
 \M$.  This observer is stationary if there exists a Killing field
 $X^T$, for some 
 $T\in {\frak g}$, such that $X^T_p=X_p$, and such that the
 observer  moves along the
 integral curve of $X^T$ through $p$. This is the case for observers
 which measure the Boulware vacuum of the Schwarschild black hole for
 example, and also for geodesic observers in de Sitter space,
 Minkowski space\ldots  It 
 may well be that $X^T$ is not causal outside a  region around 
 $p\in \M$, thus  we cannot always talk of a family of
 equivalent observers on the $G$-orbit of $p$. For an observer
 translated  from $p$ by an amount $t$ 
 along the flow of $X^T$,  a
 physical state  ${\ket
 u}\in \F$ given in the frame at $p$ corresponds,  by what we
 called a coordinate transform,  to the following:  
\ben \label{quantum vector c.o.f.2}
  {\ket v} =e^{{- it \over  \hbar}\widehat{T}}{\ket u}
\een 
One {\it postulates} that this is in fact the time evolution of the state
 vector {\it as perceived} by the observer with trajectory
 $t\mapsto 
 \exp tT.p$ in $M$: for all $t\in \mathbb{R}$,
\ben \label{time ev of vector}
\ket{u_T(t)} \equiv  e^{{- it \over  \hbar}\widehat{T}}{\ket u}
\een 
This is the Schr\"odinger picture for stationary observers. 
The subscript $T$ in $\ket{u_T(t)}$ underlines the fact that there is a chosen time
 direction $X^T$ in $(\M, h)$. 
 Of course
 $\widehat{T}$ should be interpreted as the Hamiltonian for the given
 observer. It does not vary with the observer's time precisely because
 he is stationary. Equation (\ref{time ev of vector}) implies the Schr\"odinger-type 
 equation:
\ben \label{Schrodinger eq}
i\hbar{{\rm d} \over {\rm d}t}\ket{u_T(t)}=\widehat{T}\ket{u_T(t)}
\een 
Mathematically, this holds in fact for any Killing vector field $X^B$
 in $(\M, h)$: 
letting $\ket{u_B(b)}\equiv e^{{-ib \over \hbar}\widehat{B}}\ket{u}$
 for any $B\in {\frak g}$ and $b\in \mathbb{R}$, we have:
\ben \label{time ev of vector 2}
i\hbar{{\rm d} \over {\rm d}b}\ket{u_B(b)}=\widehat{B}\ket{u_B(b)}
\een 
If $X^B$ is time-like future directed at a point $q\in \M$, then a stationary
 observer moving along $b\mapsto \exp bB.q$ will see the quantum states
 evolve according to the Schr\"odinger-type equation
 (\ref{time ev of vector 2}), as opposed to (\ref{Schrodinger
 eq}). To check questions of relativistic invariance
 for observables and the Schr\"odinger equation, the following formula
 is most useful
 \cite[p.117]{Helgason}:
\ben \nonumber 
e^{{-it \over \hbar}\widehat{T}}e^{{-ib \over \hbar}\widehat{B}}e^{{+it \over
 \hbar}\widehat{T}}= \exp\big( e^{{-it \over \hbar}\widehat{T}}\,
(-ib/\hbar)\widehat{B}\, e^{{+it \over 
 \hbar}\widehat{T}}  \big).   
\een
Basically, properties of the Adjoint
 action of the unitary group $\pi(G)$  garentee that the dynamical laws given
 by (\ref{time ev of vector}) and (\ref{Schrodinger eq}) transform
 accordingly  under changes of
 inertial frames. Details can be found in \cite{Sudarshanetal}, where
the quantum change of frame formula (\ref{quantum c.o.f.}) is postulated
 without referring to symmetries of a spacetime.   \\
  
The definition of time evolution in (\ref{time ev of vector}) suggests
the Hilbert space $\F$ should be equipped with a causal structure
depending on the choice of a particular observer. 
For each fixed spacetime point $p\in \M$, $\F$
 is endowed with a cone of infinitesimal operators which determine
 the possible
 unitary time evolutions for stationary observers at that
 point. Indeed, 
 recalling that $C^p_e\equiv  \{ B\in {\frak
 g}/ X^B_p\in C_p\}$, 
\ben 
\widehat{C^p_e}= \{ \widehat{B} \in {\rm Herm}(\F) \, / \,B\in {\frak
 g}, \, X^B_p\in C_p \;\} \label{cone of Hamiltonians at p}
\een
defines a cone of observables. Physically, $\widehat{C^p_e}$ is the
set of possible Hamiltonians for
 stationary observers at $p\in \M$. 
If the symmetry group $G$ induces causally preserving diffeomorphisms
 of the Segal structure $(\M,C_p)$, we have:
\ben
Ad_{\pi(g)}\big(\widehat{C^p_e}\big)=\widehat{C_e^{g.p}}
\een
As a consequence, if $\widehat{H}$ is a Hamiltonian for a stationary
 observer at $p\in\M$, then
 $e^{{-ib \over \hbar}\widehat{B}}\widehat{H} e^{{+ib \over
 \hbar}\widehat{B}}$ is a Hamiltonian for a stationary observer at
 $\exp bB.p$. 
\begin{Lem} \label{quantum global time lem} Let $\M$ a spacetime with Segal
 structure $(\M, C_p)$, and let $G$ act on $\M$. Let
 $\widehat{\phantom{U}}:\frak g\to {\rm Herm}(\F)$ the Dirac map of a
 theory with symmetry group $G$. Then
 the Hilbert space $\F$ admits a global time-parameter valid for particular
 observers at each spacetime point if and
 only if $(\M, C_p)$ is static. 
\end{Lem}
{\it Proof:} This is equivalent to the existence of a self-adjoint
operator $\widehat{A}\in 
\widehat{{\frak g}}$ such that for
all $p\in \M$, $X^A_p\in Int(C_p)$. $\Box$ \\
This is true whatever the state space $\F$. It really suggests that the notion of observer-independence of
future-directedness we introduced with Killing fields, is
meaningful: given any two points in $\M$, two observers
at these points stationary with respect to the same
static Killing field, see states of the theory evolve under the same
Hamiltonian. This is not possible in non-static spacetimes. 
Here
again, when the $\nu_g$ are causally preserving diffeomorphisms of
$(\M, C_p)$, 
\ben \label{Hilbert cone}
\widehat{C_e}\equiv \bigcap_{p\in M}\widehat{C^p_e}
\een
defines a sub-cone of the cone of self-adjoint operators on $\F$,
which is invariant  under the Adjoint action of the
finite-dimensional unitary group $\pi (G)$. It defines a set of
``globally valid'' Hamiltonians. Lemma \ref{causal killing field lem}
and  the previous Lemma imply that such Hamiltonians cannot exist if
$G$ does not have an Adjoint invariant cone. \\

We now determine the conserved quantities and observer-independent
observables.   It is natural
to go to the Heisenberg picture. For a stationary observer moving along
the curve $t\mapsto \exp tT.p$, the observables $\widehat{A}$ 
 evolve according to the  $\widehat{A}_T(t)\equiv e^{{ it \over
 \hbar}\widehat{T}} 
 \widehat{A} \, e^{{-it \over \hbar}\widehat{T}} $, and the 
 quantum states  $\ket{u} \in \F$ are fixed in this
 picture. $\widehat{A}$ is the observable in the Schr\"odinger
 picture, and it can have explicit time dependence. Using
diagram (\ref{quantum com diag}) and the fact that
 $\widehat{T}_T(t)=\widehat{T}$, we get
\begin{align}
{{\rm d} \over {\rm d}t} \widehat{A}_T(t) &={1 \over i\hbar} [
  \widehat{A}_T(t), \widehat{T}] +  e^{{it \over \hbar}
  \widehat{T}}\,{\partial \widehat{A} \over \partial t}\, 
  e^{{-it \over \hbar}  \widehat{T}} \nonumber \\
 &= {1 \over i\hbar} e^{{it \over \hbar} \widehat{T}}\,\Big(
  [\widehat{A}, \widehat{T}] +i\hbar{\partial \widehat{A}
  \over \partial t} \Big)\,  
  e^{{-it \over \hbar}  \widehat{T}}  
\label{quantum obs evolution} \\
&= \, \widehat{\phantom{U}}\Big(Ad_{\exp -tT}\Big( [A,T]  +i\hbar{\partial A
  \over \partial t} \Big)\Big)  \nonumber 
\end{align}
We have used the linearity and smoothness of $\widehat{\phantom{U}}$
to write $\partial \widehat{A}/\partial t$ as
$\widehat{\phantom{U}}(\partial A /\partial t)$. The first line remains
valid for any observable in ${\rm Herm}(\F)$. If 
 $\partial \widehat{A}/ \partial t=0$,  $\widehat{A}_T(t)$ defines a
constant of motion for the observer with Hamiltonian $\widehat{T}$ if
an only if $[\widehat{A}, \widehat{T}]=0$. The   
elements of  
\ben  \nonumber 
z(T)\equiv \{ A\in \U({\frak g}_{\mathbb{C}})\,/\, A^{\ast}=-A, \;
[A,T]=0\,\}
\een  
satisfy these conditions. The fact that
these depend  upon the
choice of future direction $T$ is not surprising. For example, in
Minkowski space, say there is a physical system with constant angular momentum in
the $(x,y)$ 
plane for an observer with future $\partial /\partial t$. The
observable associated to $L_z=x\partial /\partial y - y\partial /\partial x$  will satisfy 
$[\widehat{T}, \widehat{L_z}]=0$. For a different observer with future $\partial/\partial t
+(1/2)\partial/ \partial x$, we have $[\widehat{T}+(1/2)\widehat{X},
  \widehat{L_z}]=(1/2)\widehat{Y}$ which may be non-zero. Then the
observable $\widehat{L_z}$ (and its expectation value) is not
conserved for this observer, but this
is expected since he moves along the $x$-direction.

\begin{Def} A quantum  observable 
$\widehat{A}\in {\rm Herm}(\F)$ is
  called observer-independent if, 
for any spacetime inertial transform $\nu_{\exp bB}$ and any physical
  state ${\ket u}\in \mathcal{F}$, we have:
\ben \label{invariant quantum obs}
\bra{u} e^{{- ib \over \hbar}\widehat{B}} 
 \widehat{A} \, e^{{ib \over
     \hbar}\widehat{B}}\ket{u}=\bra{u}\widehat{A}\ket{u} 
\een 
\end{Def}
In theories where the vacuum state $\ket{0}$ is defined to be
invariant under inertial transforms, i.e. $\ket{0}=e^{-{ib \over
    \hbar}\widehat{B}}\ket{0}$, the vaccum expectation values of all
 the observables are invariant under change of observer.
The Casimirs of a Lie algebra are defined as the elements of the
center of its
universal algebra.  
\begin{Lem} \label{invariant quantum obs lem} Let $\M$ a
  spacetime and $G$ a Lie group acting on $\M$. Let $\widehat{\phantom{U}}$ 
 the Dirac map of a theory with symmetry $G$. The observer-independent
  observables are the elements in
\ben
\{ \widehat{A}\in {\rm Herm}(\F) \, / \, \forall \widehat{B}\in
   \widehat{{\frak g}}, \; [\widehat{A},\widehat{B}]=0 \},
\een
so that the anti-self-conjugate Casimirs of the Lie
  algebra ${\frak g}_{\mathbb{C}}$ define such 
  observables. The latter are in one-to-one
  correspondence with the Casimirs of ${\frak g}$, and their
  associated expectation values are called quantum numbers.  
\end{Lem}  
{\it Proof}: Let $\widehat{A}\in {\rm Herm}(\F)$, and suppose that for any
    $\widehat{bB}\in \widehat{{\frak 
    g}}$, (\ref{invariant quantum obs}) is satisfied. Since the scalar
    product on  $\mathcal{F}$ is
non-degenerate and $\widehat{A}$ is self-adjoint, this is equivalent
    to  $\widehat{A}= e^{{- ib \over \hbar}\widehat{B}} 
 \widehat{A} \, e^{{ib \over \hbar}\widehat{B}}$. Using 
 (\ref{quantum obs evolution}) --replacing $t\widehat{T}$ by
 $b\widehat{B}$-- it is equivalent to $[\widehat{A},
   \widehat{B}]=0$. This is satisfied when $A$ is a Casimir of
    ${\frak g}_{\mathbb{C}}$ and  $A^{\ast}=-A$. \\
One can show easily that both the self-conjugate and
    anti-self-conjugate part of a Casimir must be a Casimir. Since
    this decomposition is unique, the map (\ref{anti-self-conjugate
    isomorphism})  shows that the anti-self-conjugate part of
the center of $\U({\frak g}_{\mathbb{C}})$  is isomorphic, as a vector
    space, to the center of $\U({\frak g})$.  $\Box$ \\
Lemma \ref{invariant quantum obs lem} is the quantum
version of Lemma  \ref{obs indep
  functional lem} which could be generalised any classical observables
    in $\U({\frak g})$ or ${\rm Herm}(\U({\frak
    g}_{\mathbb{C}}))$. For example,  all simple
Lie groups admit a quadradic
Casimir constructed from the Killing form \cite[chapter 14]{FuchsJ},
whereas they  do not possess
non-trivial 
Adjoint-invariant vectors. 

It is a well-known fact in  that the Casimirs help to classify the
irreducible representations of Lie algebras, and that they are
interpreted as particular quantum numbers of a physical system with
symmetry group $G$. The language of observer-dependence is meaningful:
typically, angular 
momentum about an axis can be observer-dependent, whereas spin is
not. Using equation (\ref{quantum obs
  evolution}), the most important physical consequence of
 observer-independence, is that the Casimirs yield constants
of motion for {\it all} inequivalent stationary
observers of spacetime, {\it whatever their
  respective Hamiltonians.}

\subsubsection{Inequivalent Hamiltonians}

We have not mentioned physical issues relating to the existence
of a set of normalized eigen-vectors for the
quantum observables. Finding basis of eigen-vectors of
particular observables, especially the Hamiltonian, is an essential
step towards understanding the physics. Proposition \ref{quantum univ
  prop} shows that Hamiltonians related by a change of inertial frame
(\ref{quantum c.o.f.}) trivially share the same spectrum and their
eigen-states are related by a unitary transform. However, physically
inequivalent Hamiltonians will generically have different
properties, especially 
regarding normalizability of the eigen-states, and boundedness of the
spectrum.  Stationary observers do not generically have equivalent
futures, even 
for geodesic observers. This does not
contradict  the principle of relativity
of course, which holds for {\it observers related by an inertial
transform}. Consider  for
example a Rindler observer in Minkowski space moving on the
time-like 
integral curve of the Killing field $z\partial_t +t\partial_z$ through
$(t=0,z=1/a)$. The
Minkowski vacuum $\ket{0}$ is defined with respect to $\partial_t$
which defines  
a ``global time'' on the Hilbert space,  as in Lemma
\ref{quantum global time lem}.  These two Killing fields are not
inertially related since the former has a horizon while the latter does
not. If the Dirac map is injective, their associated Hamiltonians
cannot be Adjoint related and thus are physically inequivalent. Indeed, the Rindler
observer, who  
moves with uniform acceleration $a$, perceives $\ket{0}$ not as
the lowest energy eigen-state of his Hamiltonian, but as 
a thermal bath at temperature $a/2\pi$
\cite{Unruh}. If the observers were  equivalent,
the vacuum  $\ket{0}$, since it is invariant the unitary action of
$\pi(G)$, would 
be an energy eigen-state for the Rindler observer, which it is not.\\
Our approach does not explain how to obtain this particular result, but
gives a general frame to study and understand the reasons for 
observer-dependence in theories
with symmetries acting on a  spacetime: 
\begin{Prop} \label{inequivalent hamiltonian prop} Let $(\M, h)$ a time-orientable spacetime,  $(\M, C_p)$
  its associated Segal structure, and $G$ its connected group
  of Killing  motions. Consider a quantum theory on
$(\M,h)$ with symmetry group $G$, and call $\pi$ the associated representation 
of $G$. Suppose that the derived representation of ${\frak
  g}$ is faithful, and call $H_p$ the stabiliser subgroup of $p\in
  \M$. Then the set of physically inequivalent Hamiltonians for
  stationary observers at $p\in \M$, is in one-to-one correspondence
  with the projective orbits of the cone
\ben
{C_e^p}\equiv \{A\in {\frak g} \, / \, X^A_p\in C_p \;\} \label{cone
  and stationarity}
\een
under the Adjoint action of $H_p$ in ${\frak g}$. Furthermore this
  classification  is invariant under changes of inertial frames, so
  all the inequivalent  Hamiltonians of the quantum theory on
$(\M,h)$, are obtained by taking one such classification on each 
$G$-orbit in $\M$.      
\end{Prop}
{\it Proof:} By Lemma \ref{connected Lie actions}, the action of $G$ on
$(\M, C_p)$ is
causally preserving. The map $A\in {\frak g} \mapsto X^A$ is injective,
and, because of coordinate reparametrisation, each ray  in $C^p_e$
defines a stationary observer at $p\in \M$ and reciprocally. Call
$\mathbb{P}(C^p_e)$  the set of rays in
$C^p_e$. Since $\nu_h(p)=p$ for $h\in H_p$ and the action is causally
preserving, it follows from
(\ref{adjoint related cones 3}) that $Ad_h(C^p_e)=C^p_e$ and one can
define  $\mathbb{P}(C^p_e)/Ad_{H_p}$. Clearly two elements $A,B\in
\mathbb{P}(C^e_p)/Ad_{H_p}$ define the same equivalence class if and
only if there exists $h\in H_p$ such that $X^A= \nu_h\ast
X^B=X^{Ad_hB}$, and $\mathbb{P}(C^p_e)/Ad_{H_p}$ defines the set
of inequivalent stationary observers at $p\in \M$.  \\
Now  $\widehat{C^p_e}$ as in (\ref{cone of Hamiltonians at p}) is
the set of Hamiltonians for these stationary observers, and when
$\pi\ast$ is faithful it is isomophic  to
$C^p_e$. It follows directly from Proposition
\ref{quantum univ prop} that elements of  $\mathbb{P}(C^p_e)/Ad_{H_p}$ label
mutually inequivalent Hamiltonians. \\
The fact that it suffices to consider only one point on each $G$-orbit
simply follows from the nature of the changes of inertial
frames. Formally, since 
$H_{g.p}=gH_p g^{-1}$ and $Ad_g(C^p_e)= C^{g.p}_e$, the map:
\ben \nonumber
Ad_g:\mathbb{P}(C^p_e)/Ad_{H_p}\longrightarrow
\mathbb{P}(C^{g.p}_e)/Ad_{H_{g.p}} 
\een
defines a bijection between the equivalence classes of stationary
observers at $p$ 
and at $g.p$. The corresponding Hamiltonians are related according to 
$\widehat{A}\mapsto \pi(g)\widehat{A}\pi(g)^{-1}$, so that their
physical properties are the same.  $\Box$ \\
Note that this is substancially different from the classification of
inequivalent time-like geodesics at a point $p\in\M$. Indeed, the latter only
requires finding the orbits of the tangent vectors in $C_p$ under the
linear-isotropy 
group $\nu_{H_p}$, whereas here one must consider the Killing 
  fields.  \\
For a fixed
point $p\in\M$, $C^p_e$ defines a cone in ${\frak g}$, and thus a
corresponding left-invariant Segal structure $(G,
L_g\ast(e)C^p_e)$. Proposition  \ref{inequivalent hamiltonian prop}
tells us that classifying the inequivalent Hamiltonians of the theory
for an observer at $p$,  amounts to classifying
one-parameter subgroups  of $G$ which are causal with respect to the
Segal  structure $(G,L_g\ast(e)C^p_e)$, under the action of $Ad_{H_p}$.\\

We can classify the inequivalent Hamiltonians for
observers in $AdS_2$, using the results and notation 
 of  Sec.\ref{AdS2 example}. The global, 
Poincar\'e, and Schwarzschild Killing fields correspond respectively to
timelike, null and spacelike elements in ${\frak sl}(2,\mathbb{R})$
equipped with its Killing form. In the
$H,K,D$  basis of ${\frak sl}(2, \mathbb{R})$ defined by:
\begin{align*}
\sqrt{2}T\equiv{1 \over 2}(H+K), \quad \sqrt{2}Y\equiv {1
  \over2}(H-K), \quad \sqrt{2}Z\equiv D,
\end{align*}
the Killing fields read:
\begin{align*}
{\partial \over \partial \tau}=X^{{1 \over 2}(H+K)}, \quad {\partial
  \over \partial x}= X^H, \quad {\partial \over \partial q}=X^{{1
  \over 2}(H-K)}.
\end{align*}
The
property of their respective Hamiltonians vary according to the
theory. In quantum field theory on $AdS_2$, the global and Poincar\'e vacua are
unitarily related, but differ from the Schwarzschild vacuum
\cite{Strom2}; this should be manifest in the lowest eigen-states of
the Hamiltonians 
$\widehat{H}+\widehat{K}$, $\widehat{H}$ and
$\widehat{H}-\widehat{K}$, but subtleties of convergence may arise as
the stationary observers for $H$ and $H-K$ may move on light-like or
spacelike curves in $AdS_2$, according to the point chosen. \\   
Let us  consider Maxwell-Einstein theory on $AdS_2\times S^2$,
and  focus on the non-relativistic motion of particles with
charge equal to 
mass. This corresponds to the motion of certain charged particles in
the infinite 
mass limit of the near-horizon geometry of the extremal
Reissner-Nordstr\"om black hole. It was shown in 
\cite{ClausandDerix} that the dynamics is in fact equivalent to
the one-particle conformal mechanics model studied
by de Alfaro, Fubini and Furlan (DFF)in \cite{deAlfaro}. In spite of the
large $R$ 
limit taken to obtain these dynamics, the quantum operators
$\widehat{H}, \widehat{K}, \widehat{D}$ stemming from the phase-space
functions  of the DFF model define a 
representation of ${\frak sl}(2,
\mathbb{R})$, so that the relativistic symmetries of $AdS_2$
correspond to conformal  
symmetries of
this non-relativistic theory. From \cite{deAlfaro}, the
Hamiltonian 
$\widehat{H}+\widehat{K}$ 
 has an evenly spaced discrete spectrum with
normalizable eigen-states, whereas 
 $\widehat{H}$ 
has a continuous spectrum bounded from below, and its lowest
eigen-state is not normalizable. These operators govern time
evolution respectively for    
global stationary observers or Poincar\'e stationary
observers. Mathematically at least,
their states 
 evolve in a {\it different} Hilbert space: the inequivalence of
the  Hamiltonians becomes all the more physically relevant. \\
We now apply Proposition \ref{inequivalent hamiltonian prop} to find
all the inequivalent 
Hamiltonians at the point $p$ with $(\tau=\pi/2,\sigma=\pi/2)$ in
the global coordinates. Since $AdS_2$ is homogeneous, it suffices to
look at this point. We have:
\ben \nonumber
C^p_e=\{tT+yY+zZ\,/\,t\geq 0, \,t^2\geq y^2 \,\}   
\een
and $H_p$ is the one-parameter subgroup generated by $Z$. Its Adjoint
action on $C^p_e$ is equivalent to  a Lorentz velocity boost in the
$(t,y)$ plane of ${\frak sl}(2, \mathbb{R})$. Thus for each fixed
$z\in \mathbb{R}$, there are three projective orbits  labelled by a
strictly timelike element such as
$H+K+zZ$, a null element such as $K+zZ$, and a null element such as $H+zZ$.   
The Hamiltonian $\widehat{H}$ here corresponds to a stationary observer moving
at the speed of light (i.e. on the horizon of $\partial_x$), but at
other points it corresponds to strictly-timelike observers. 

\subsection{Highest weight representations}
We now explain how Adjoint invariant cones 
 in ${\frak g}$ are related to sets of positive operators on $\F$. In Lemma
 \ref{quantum global time lem} we established that such cones were
 necessary in order to define of a
 Hamiltonian which is globally valid for particular observers in
all of $\M$. The assumptions are simply that the space of physical states of
 the theory carries a unitary
representation of a symmetry group which acts on the spacetime,
 thereby defining inertial transforms. Recall that this is the case in
 quantum field theories, and also in string
theories: although one tends to concentrate on the string world-sheet Virasoro
symmetries, the target space or global symmetries, whether one works
 in the light-cone gauge or not, are essential. The
 physical spaces $\F$ admit a bounded
notion of energy whenever there exists a self-adjoint operator
which has a (one-side) bounded spectrum and which can be thought of as
generating 
time evolution. In the Schr\"odinger picture for stationary observers
 (when possible),
we would want this  
operator to be the image by the Dirac map of a Killing field which is
causal for a particular observer in spacetime. \\
Representations which admit one-side bounded operators 
are in fact  unitary highest weight representations \cite[Theorem
 X.3.9]{Neeb}. They possess rich
algebraic and geometric structures, whose mathematical properties can
 be found in 
\cite[Part D]{Neeb}. From a physical point of view, the algebraic
 structure  essentially  enables
 the construction of Fock spaces by acting on a highest weight vector,
 while the  geometric structure encodes the existence of
positive energy operators. It has been known for some time that the de-Sitter
group $O(1,4)$ does not admit such a representation \cite{Wigner3}. In
fact as we see now, the same is  true for all groups with ambivalent
conjugacy classes. \\
As in the previous section, let $G$ a  Lie group, $\pi$ a continuous
 unitary representation of 
 $G$ on a Hilbert space
$\F$, and  $\pi \ast$  the corresponding derived 
 representation of ${\frak g}$ as defined in (\ref{derived repr}). 
 Strictly speaking, $\pi\ast$ is defined
on a dense subset of $\F$ consisting of smooth vectors. For all $A\in
 {\frak g}$  and $h \in G$, we
 have:
\begin{align}
 \pi \ast (Ad_hA) &=  Ad_{\pi(h)} \pi \ast (A) \nonumber  \\
\label{ambivalent}                  &= \pi(h)\pi \ast(A) \pi(h^{-1}),
\end{align} 
which is similar to equation (\ref{quantum c.o.f.}). If $G$ has ambivalent
conjugacy classes, then for all $A\in {\frak g}$, 
there exists $h\in G$
such that:
\ben \label{ambivalent adjoint}
 Ad_h A= -A.\een
Since  $i\pi \ast (A)$ is a self-adjoint operator, let 
 $\ket{v} \in
\F$ and $\lambda \in \mathbb{R}$ such that $i\pi \ast (A)\ket{v}=\lambda
\ket{v}$. Then using (\ref{ambivalent adjoint}) and
(\ref{ambivalent}), we get:  
\begin{align*}
i\pi \ast (Ad_hA) \pi(h) \ket{v} &= -i\pi \ast (A) \pi(h) \ket{v} \\
                      & = i\pi(h)  \pi \ast(A) \pi(h^{-1}) \pi (h) \ket{v} \\
                    & = \lambda i \pi(h)\ket{v}
\end{align*} 
Thus $\pi(h)\ket{v}$ is an eigen-vector of $\widehat{A}$ with eigen-value
$-\lambda$. The spectrum of the observable $\widehat{A}$ is
symmetric about 0. \\
Now suppose $G$ acts on a
spacetime $\M$ by inducing causally preserving diffeomorphisms of a
Segal structure $(\M, C_p)$. The operator $\widehat{A}$ defines a possible Hamiltonian for
an observer at $p\in \M$, if $X^A_p\in Int(C_p)$.  For such an observer
the states $\ket{v}$
and  $\pi(h)\ket{v}$ have respective energies
$\lambda=\bra{v}\widehat{A}\ket{v}$  and
$-\lambda=\bra{v} \pi(h)^{-1} 
 \widehat{A} \pi(h)\,\ket{v}$. By definition  of
changes of observers, $-\lambda$ is also the energy
measured for the state $\ket{v}$ by an equivalent observer at
$h.p$ with same Hamiltonian $\widehat{A}$. This observable 
corresponds to the ``future'' $X^{A}_{h.p}$ at $h.p$. However, since
$\nu_h$ is a causally preserving diffeomorphism, using (\ref{adjoint c
  o f}) and 
(\ref{ambivalent adjoint}) we have:
\ben \label{necessary horizon}
\nu_h\ast(p)X^A_p=X^{Ad_hA}_{h.p}=-X^A_{h.p} \in C_{h.p}
\een
so that in fact $X^A_{h.p}\notin C_{h.p}$.  The observer at $h.p$ with
Hamiltonian $\widehat{A}$ is in fact
measuring minus the energy in his frame. Of course the inertially
equivalent observer on the other hand
measures $\pi(h)\ket{v}$ with $\pi(h)\widehat{A}\pi(h)^{-1}$ and gets
the same expectation value $\lambda$. More importantly,
(\ref{necessary horizon}) 
implies that the two
observers measuring the opposite energy for the given state, are
necessarily separated by a horizon in the spacetime. From the first
observer's point of view, we could say that whereas $\ket{v}$ is
measurable in his frame, $\pi(h)\ket{v}$ lies behind his horizon. The
question is then whether one can always construct localized states
\cite{NewtonandWigner2} for theories with symmetry groups with
ambivalent conjugacy classes, or just with groups which do not admit unitary
highest weight representations.   \\

With ${\rm Herm}(\F)$ denoting the set of not necessarily bounded
self-adjoint operators on $\F$, we define:
\ben
{\rm Herm}^+(\F)\equiv \{ O\in {\rm Herm}(\F)\,/\,\forall v\in \F,\;
\bra{v}O\ket{v}\geq 0 \, \}
\een
The positive operators define  a (non-trivial) pointed closed
convex cone in ${\rm
  Herm}(\F)$.  
\begin{Lem} Let $G$ the symmetry group of a theory, 
 and  $\pi$ a
unitary  representation of $G$  on the Hilbert space
 $\F$ of this theory, such that the derived representation $\pi\ast$
of ${\frak g}$ is faithful. If there
  exists a non-trivial positive  
 observable $i\pi\ast(A) \in {\rm Herm}^+(\F)$ for some $A\in {\frak
   g}$, then $G$  admits a non-trivial 
 pointed $Ad_G$-invariant closed convex cone. 
\end{Lem}
{\it Proof:} ${\rm Herm}^+(\F)$ is
invariant under conjugation by unitary transforms of $\F$, so in
particular for 
all $g\in G$, $\pi(g){\rm Herm}^+(\F)\pi(g^{-1})\subset{\rm
  Herm}^+(\F)$.  The same is true for $i\pi\ast({\frak g})$, so  
$i\pi\ast({\frak g})\cap {\rm Herm}^+(\F)$  is an
$Ad_{\pi(G)}$-invariant pointed convex cone. By
hypothesis it is non-empty. Its closure is in ${\rm Herm}^+(\F)$ hence
it is also
$Ad_{\pi(G)}$-invariant and pointed. The 
inverse image of the latter by $\pi\ast$ defines a cone in ${\frak g}$
with the necessary requirements.
It is pointed when $\pi\ast$ is injective, and does not contain $-A$. $\Box$ \\
This simple property is completely analogous to Lemma \ref{causal killing
  field lem}: causal Killing fields
on spacetimes or  positive
observables associated  to
symmetry generators in any quantum theory, require the existence of
Adjoint invariant 
cones in the symmetry groups.  In fact the
study of one-side bounded operators --rather than positive
operators--, is more subtle, and requires
introducing mathematical tools which go beyond the present aim of this
article.

\begin{Thm} \label{bof bof}
Let $G$ a non-compact semisimple Lie
  group with maximal compact subgroup $K$. Then $G$  admits
  unitary highest weight representations if and only if $G/K$ is a
  hermitian symmetric space.   
\end{Thm} 
{\it Proof:} This is a consequence of Theorem IX.5.13 in \cite{Neeb}
which is more general and relates unitary highest weight representations of involutive Lie algebra to the existence of invariant generating convex
sets. $\Box$ \\
Some
properties of unitary
highest weight representations of these groups can be found in
\cite{Gunaydin}: the 3-grading of the Lie algebras
 of the groups $G$ in Table \ref{irred herm} p.\pageref{irred
  herm} implies one can construct Fock spaces by acting with creation
operators on  highest weight vectors. \\  
Together with Theorem \ref{Segal thm}, Theorem \ref{bof bof} states the
equivalence between the existence of positive energy representations
for semi-simple 
groups and that of bi-invariant pointed Einsteinian cones in their Lie
algebras. These are related to causal Killing fields on manifolds, and
we get the following:
\begin{Thm}\label{ULWR thm} Let $G$ a simple non-compact Lie group  act almost
  effectively on a spacetime $\M$, and suppose that the $\nu_g$, for all $g\in
  G$, are causally preserving diffeomorphisms of a Segal structure
  $(\M, C_p)$. If $(\M, C_p)$ admits a causal Killing field, then
  there exists a unitary highest  
  weight representation of $G$, so that one can define through the
  Dirac map a quantum theory  which has a notion of positive energy.
\end{Thm}
{\it Proof:} From Lemma \ref{causal killing field lem}, if $(\M, C_p)$
admits a causal Killing field,  then
${\frak g}$ admits an $Ad_G$-invariant (non-trivial pointed closed
convex) cone , which is Einsteinian for $G$ simple. Theorem \ref{Segal
thm} implies that $G/K$ is a non-compact
hermitian symmetric space, and we apply Theorem \ref{bof bof}. $\Box$
\\
This theorem applies of course to Lorentzian spacetimes with static
metrics and non-compact simple Lie groups. We see  
that the classical definition of energy, the existence of
bi-invariant cones in the group of motion, and that
of a bounded quantum energy, are inherently intertwined.
Theorem \ref{ULWR thm}
and  Lemma \ref{quantum global time lem}, imply that, for
spacetimes with non-compact simple symmetry groups, 
if one requires the existence of a global time in quantum theories,
then one can also choose a 
unitary highest weight representation, and thus have positive energy.
\\

Our analysis of symmetries in quantum theories relies on promoting
the infinitesimal generators to self-adjoint quantum observables.  
The discussion in Sec.\ref{Dirac procedure section} on the Dirac procedure
applied to the universal algebra, 
is relevant to any theory which
requires at some point  a
unitary representation of a symmetry group, whether compact or
non-compact, relativistic or non-relativistic. The comments on time
evolution and the causal
structure of the Hilbert space rely however on the existence of
stationary 
observers.  
By definition this is the case in all spacetimes with $G$-orbits which are
 non-spacelike, thus in all homogeneous spacetimes of course, and most
 probably in all backgrounds where 
string theory is tractable. A different approach regarding
time-evolution must be taken in  the cosmological models with spatial
symmetry only.  For non-stationary observers, even in homogeneous spacetimes,
time-evolution will require
 time-dependent Hamiltonians, and the energy for such observers will 
 not generically be conserved. \\
The symmetry generators in the Dirac procedure are analogous to the
``fundamental quantities'' in Dirac's paper \cite{Dirac1}. Owing
to the explicit group structure in the
formalism given here, the
observer-dependent 
choice of Hamiltonian does not break relativistic invariance, and at
this formal stage, there is no need to introduce coordinate charts.    
However, the quantum observables described in this procedure  
are only those  which correspond to
the generalised Killing  classical observables of Proposition \ref{universal
  algebra prop}. Although these are particularly easy to manipulate in
terms of changes of frames and observers,
they cannot not exhaust all the observables, especially when the spacetime
is not homogeneous. More fundamentally, the universal algebra only 
contains differential observables, and thus does not contain the
analogue of position in phase-space. Mathematically, if $1$ denotes 
the unit element of $\U({\frak g}_{\mathbb{C}})$, for any two elements  $A,B\in
\U({\frak g}_{\mathbb{C}})$, we cannot have 
$AB-BA= 1$  \cite[Remark
  2.8.3]{Dixmier}: the 
commutator of any two quantum observables which classically correspond
to elements 
of $\U({\frak g}_{\mathbb{C}})$, cannot be proportional to the
identity operator 
on $\F$. It can be central though, but to obtain standard first quantization
 commutation relations, one has to extend $\U({\frak
  g}_{\mathbb{C}})$.

\section{Conclusion}
\label{conclusion}
The main achievement of this article is to explain the physical
relevance of bi-invariant cones 
in symmetry groups of spacetimes, and show how they appear naturally
in most supergravity and string theories currently under
investigation.   

In order to interpret Adjoint invariance physically, 
we gave a detailed description of the
way  symmetries act on physical theories, insisting especially
on the mathematical relations between the symmetries of a space
of states, and the symmetries of a spacetime. Propositions
\ref{universal algebra prop} and \ref{quantum univ prop} show how the
universal algebra of the infinitesimal symmetries both corresponds to
spacetime observables --differential operators on $\M$--, and quantum
observables --self-adjoint operators on $\F$--, and that changes of
inertial frames in $\M$ correspond to Adjoint actions on these
observables. The notions of changes of inertial frames and changes of
observers were clearly distinguished. \\
Most importantly, we showed that Adjoint invariance is fundamentally related to
invariance under changes of observers in a theory. For example, Lemmas \ref{obs
  indep functional lem} and \ref{invariant quantum obs lem} show that
the Casimirs  of
the Lie algebra of symmetries define observer-independent
observables. The key novel results however are those which
relate 
observer-independence of future-directedness in a spacetime, to
algebraic properties of the group of motions. In Sec.\ref{adjoint action
  and causally preserving maps}, we established the equivalence
between the existence of particular bi-invariant cones in the algebra,
and observer-independent or static
causal structures. This shows that in a spacetime, the local causal
structure defined by the future cones, together with the symmetry
group of the metric,
suffice to determine whether the spacetime admits a causal Killing
field.  These results imply that Segal's
assumption of bi-invariant cones is satisfied for all symmetry groups
of spacetimes which admit at least one causal Killing field. As a
consequence, the
theory of causal symmetric spaces  \cite{Olafsson} should have
applications in modern theoretical physics.

Furthermore, infinitesimal causal structures  were shown to be
useful to understand time evolution in quantum
theories: in Sec.\ref{Dirac procedure section} we gave a formal
relation  between 
local relativistic times, and particular quantum operators interpreted as
the Hamiltonians for stationary observers. Lemma \ref{quantum
  global time lem} shows the intuitive fact that spacetime staticity
is equivalent  to the
existence, in any quantum theory, of a Hamiltonian valid for a
family of stationary observers  at
each spacetime point, or indeed of a global time evolution on the
space of states. \\
The notion of changes of observers enabled us to
formally classify  the
inequivalent Hamiltonians of a quantum theory, by simply studying the
orbits of the locally causal Killing vectors at a point under the action of the linear isotropy
group at this point. The technique explained in Proposition
\ref{inequivalent hamiltonian prop}
relates to the classification  of quotients
of spacetimes admitting symmetries. Over the past years, this has been
widely studied in
the context of Kaluza-Klein and orbifold reductions (see \cite{Simon3}
and references therein). However, it should be noted that whereas in
these contexts one classifies the symmetries up to the Adjoint action
of the full symmetry group, here one can only use the action of the stabiliser
groups, since the Killing fields are not necessarily static, and hence
may not always define globally valid Hamiltonians.

We also showed that bi-invariant cones in symmetry groups encode the
possibility  of
defining positive energy functionals. Lemma \ref{causal killing field
  lem} shows their nessecity in classical theories, while Theorem
\ref{bof bof} shows their necessity and sufficiency for quantum
theories with certain symmetry groups. As we said, more general
results relating bi-invariant cones to unitary highest weight
representations can be found in \cite[Part D]{Neeb}. This approach
yields a geometrical description of the set positive energy operators
as corresponding to an Adjoint invariant cone in the symmetry
group.  The fact that
symmetries play a major r\^ole in the problem of
positive energy is not new, but our achievement here is relate it to
infinitesimal causal structures: Theorem \ref{ULWR thm} shows that
for non-compact simple symmetry groups, spacetime staticity implies
the existence of positive energy representations. This should have
applications in understanding better the
physics of unitary highest weight representations,  which commonly arise, for
example, in supergravity theories \cite{Gunaydin}. One should
not forget that positive observable-functionals remain positive under
changes of observers, and such observables seem less necessary
whenever there
is a notion of localised states in the theory. The existence of
observer-independent   future-directions
relates to that of positive energy functionals simply
because observer-independence is encoded as Adjoint invariance in the
symmetry group.

\subsection*{Acknowledgements}
I am immensely grateful to Gary Gibbons for introducing me to
\cite{Segal} and suggesting its relevance to current theories. I also
thank Joachim Hilgert
and Karl-Hermann Neeb for bringing \cite{Neeb} to my notice, and
Jasbir Nagi for useful discussions on universal algebras. This work was
supported by  the EPSRC, the DAMTP, and the Cambridge
European Trust.

\rhead[\fancyplain{}{\footnotesize References}]{\fancyplain{}{\thepage}}

\bibliography{refs}
\bibliographystyle{JHEP}

\end{document}